\documentclass{amsart}

\usepackage[latin1]{inputenc}
\usepackage{amssymb}
\usepackage{amsmath}
\usepackage{epsfig}

\newtheorem{theorem}{Theorem}[section]
\newtheorem{rem}{Remark}[section]
\newtheorem{ex}{Example}[section]
\newtheorem{prop}{Proposition}[section]
\newtheorem{lemma}{Lemma}[section]

\newtheorem{defi}{Definition}[section]

\newcommand{\ZZZ}{\mathbb{Z}_3}
\newcommand{\lra}{\longrightarrow}

\newcommand{\lmt}{\longmapsto}

\newcommand{\ra}{\rightarrow}
\newcommand{\Ra}{\Rightarrow}
\newcommand{\eprf}{\hfill $\square$ \smallskip\par}

\newcommand{\PP}{ \mathbb{P}}
\newcommand{\C }{ \mathbb{C}}

\newcommand{\Z}{\mathbb{Z}}
\newcommand{\Q}{\mathbb{Q}}

\newcommand{\EEE}{E^3}
\newcommand{\phip}{\varphi_3}
\newcommand{\exijk}{\widetilde{B_{i,j,k}}}

\makeatletter
\def\blfootnote{\xdef\@thefnmark{}\@footnotetext}

\begin{document}

\title{A rigid Calabi--Yau 3-fold}

\author{Sara Angela Filippini and Alice Garbagnati}

\address{Sara Angela Filippini, Dipartimento di Fisica e Matematica, Universit\`a dell'Insubria, 
 via Valleggio 11, I-22100 Como, Italy}

\email{saraangela.filippini@uninsubria.it}

\address{Alice Garbagnati, Dipartimento di Matematica, Universit\`a di Milano,
  via Saldini 50, I-20133 Milano, Italy}

\email{alice.garbagnati@unimi.it}
\urladdr{http://sites.google.com/site/alicegarbagnati/home}

\begin{abstract} The aim of this paper is to analyze some geometric properties of the rigid Calabi--Yau threefold $\mathcal{Z}$ obtained by a quotient of $\EEE$, where $E$ is a specific elliptic curve. We describe the cohomology of $\mathcal{Z}$ and give a simple formula for the trilinear form on $Pic(\mathcal{Z})$. We describe some projective models of $\mathcal{Z}$ and relate these to its generalized mirror. A smoothing of a singular model is a Calabi--Yau threefold with small Hodge numbers which was not known before.
\end{abstract}

\maketitle
\section{Introduction}
One of the most exciting mathematical implications of string theory is mirror symmetry, which finds its origin in the papers \cite{Dixon:1987bg} and \cite{Lerche:1989uy}. A phenomenological verification of the conjecture that Calabi--Yau manifolds should appear in pairs was given in \cite{Candelas:1989hd}, and the first non trivial examples of mirror pairs appeared in \cite{Greene:1990ud}.
In \cite{Greene:1990ud} it was also discovered that mirror symmetry can be used to compute the instanton corrections to the Yukawa couplings (the first explicit computations were carried out in  \cite{Candelas:1990rm}), which mathematically corresponds to determine the number of rational curves of given degree embedded in the Calabi--Yau manifold.  This led to the notion of Gromov-Witten invariants and more generally to the one of Gopakumar-Vafa invariants \cite{Gopakumar:1998ii}, \cite{Gopakumar:1998jq}.\\
Curiously, in the seminal paper \cite{Candelas:1985en}, where the relevance of Calabi--Yau manifolds in string theory was established, among the few explicit known examples of Calabi--Yau manifolds there was the manifold $\mathcal{Z}$, realized as the desingularization of the quotient $\EEE/\phip$,  with $\phip=\varphi\times \varphi\times \varphi$ and $\varphi$ the generator of $\ZZZ$ which acts on the elliptic curve $E$. As $\mathcal{Z}$ is a rigid manifold, it cannot admit a Calabi--Yau threefold as mirror partner. This created a puzzle in the general framework of mirror symmetry. However, physically, mirror symmetry arises as a complete equivalence between conformal field theories. In this respect, it should not be surprising that in certain exceptional cases the equivalence could involve more general spaces. Indeed, in \cite{CDP} it was proposed that the mirror of $\mathcal{Z}$ should be a cubic in $\mathbb P^8$ quotiented by a suitable finite group.
By using the usual mirror methods, the authors were able to reproduce the right Yukawa couplings of $\mathcal{Z}$. The mirror symmetry generalized to rigid Calabi--Yau manifolds has been considered also in \cite{Schimmrigk:1994ya}, where the mirror is presented as (embedded in) an higher dimensional Fano variety having the mirror diamond as an embedded sub-diamond and in \cite{AG} where it is related to toric geometry. However, a definitive understanding of the question is still open.

In this paper, as a preparation to further work on generalized mirror symmetry, we present a very detailed study of the rigid manifold $\mathcal{Z}$. Section \ref{sec: EEE widetildeEEE Z and thier cohomology} is devoted to an explicit description of the cohomology of $\mathcal{Z}$. The Hodge diamond of this threefold is very well known, but here we identify a  set of generators of $Pic(\mathcal{Z})$ made up of surfaces and a set of generators of $H^4(\mathcal{Z},\Q)$ made up of curves. Our goal is to describe the trilinear intersection form on the generators of $Pic(\mathcal{Z})$ relating it with the trilinear intersection form on $Pic(\EEE)$. Indeed the generators of $Pic(\mathcal{Z})$ are of two types: the ones coming from the generators of $Pic(\EEE)$ and the ones coming from the resolution of the singularities of $\EEE/\phip$. The intersection between two divisors of different type is zero, and the trilinear form on the divisors coming from $\EEE$ is, up to a constant, the trilinear form $Pic(\EEE)$.
For this reason it is important to give a good description of the trilinear form on $Pic(\EEE)$: in \eqref{eq: trilin on EEE} the cubic self-intersection form is given for each divisor in $Pic(\EEE)$. In Section \ref{sec: more on the trilinear form on Pic(EEE)}, Theorem \ref{theorem: from L to OmegaL}, we prove that it can be given in terms of the determinant of a matrix in $Mat_{3,3}(\mathbb{Q}[\zeta])$, $\zeta^3=1$. The locus where the determinant vanishes is a singular cubic in $\mathbb{P}^8$. We recall that the Yukawa coupling on $H^{1,1}$ is strongly related with the cup product on $H^{1,1}$ and thus with the intersection form on $Pic(\EEE)$. Moreover the locus where the Yukawa coupling  vanishes corresponds to fermion mass generation points.\\
In the second part of the paper, in Section \ref{sec: projective models Z}, we describe some projective models of $\mathcal{Z}$.  Here we will limit ourselves to make some basic observation on mirror symmetry, deferring a systematic analysis to a future paper. We give a detailed description of the images of three maps (called $m_0$, $m_1$, $m_2$) defined from $\mathcal{Z}$ to projective spaces and we relate these to earlier work. None of the maps $m_i$, $i=0,1,2$, gives an embedding. For this reason, we also prove that a certain divisor on $\mathcal{Z}$ is very ample (cf.\ Proposition \ref{prop: 2F-M}), i.e\ it defines a map $m$ such that $m(\mathcal{Z})\simeq \mathcal{Z}$.\\
The maps $m_i$, $i=0,1,2$, allow us to describe some peculiarities of $\mathcal{Z}$. The map $m_0$ is $3:1$ and it gives a model of another rigid Calabi--Yau threefold $\mathcal{Y}$, birational to $\mathcal{Z}/\ZZZ$. Moreover $m_0(\mathcal{Z})$ is contained in the Fermat cubic hypersurface in $\mathbb{P}^8$ and this could give a geometrical interpretation  of the conjectures on the generalized mirrors of the rigid Calabi--Yau 3-folds $\mathcal{Z}$ and $\mathcal{Y}$ presented in \cite{CDP} and \cite{KLS}. The map $m_1$ contracts 27 rational curves on $\mathcal{Z}$, and gives a model of $\mathcal{Z}$ embedded in $\mathbb{P}^{11}$. This model will be used in Section \ref{sec: smoothing} to obtain other Calabi--Yau 3-folds. The map $m_2$ was already defined by Kimura, \cite{Ki}, to show that there exists a birational map between $\mathcal{Z}$ and a particular complete intersection of two cubics in $\mathbb{P}^5$, called $V_{3,3}$. Several models of the variety $V_{3,3}$ were analyzed previously (cf.\ \cite{Hi}, \cite{WvG}, \cite{Me}).\\ 
We already observed that the Calabi--Yau threefold $\mathcal{Z}$ is very well known, but it can be used to construct several other Calabi--Yau 3-folds, which are not always rigid. In Section \ref{sec: other CY 3-folds} we recall constructions which produce Calabi--Yau 3-folds starting from a given one. In Section \ref{sec: smoothing} we apply one of these constructions (described in \cite{Fr}) to $\mathcal{Z}$ and we obtain non rigid Calabi--Yau threefolds. The idea is to contract some curves on $\mathcal{Z}$ and then to consider the smoothing of the singular threefold obtained. One of the Calabi--Yau threefolds constructed in this section does not appear in the list of known Calabi--Yau threefold with small Hodge numbers given in \cite{Braun:2009qy} and is a new Calabi--Yau variety.

\section{The 3-folds $\EEE$, $\widetilde{\EEE}$, $\mathcal{Z}$ and their cohomology}\label{sec: EEE widetildeEEE Z and thier cohomology}
In this note we will analyze the properties of the very well known Calabi--Yau 3-fold $\mathcal{Z}$ introduced independently in \cite[Example 2]{Beauville} and \cite{SW}. In order to describe the trilinear form on $Pic(\mathcal{Z})$ (cf.\ \eqref{formula: cubic form}), which is strongly related to the Yukawa coupling, we will compute the cohomology of $\mathcal{Z}$ (Section \ref{section: the cohomology of Z}) and of the varieties involved in its construction (Sections \ref{section: cohomology of EEE} and \ref{section: the cohomology of widetildeEEE}).\\ 
To fix the notation we recall some definitions and the construction of $\mathcal{Z}$. 
\begin{defi} A smooth compact complex variety $X$ is called a Calabi--Yau variety if it is a K\"ahler variety, it has a trivial canonical bundle and $h^{i,0}(X)=0$ for $0<i<dim(X)$.\end{defi}
To give the Hodge diamond of a Calabi--Yau 3-fold $X$ one has to find $h^{1,1}(X)$ and $h^{2,1}(X)$. We recall that $h^{2,1}(X)$ is the dimension of the family of deformations of $X$ (which are indeed unobstructed by the Tian--Todorov theorem), so $X$ has $h^{2,1}(X)$ complex moduli.\\

Let $E$ be the Fermat elliptic curve $x^3+y^3+z^3=0$, i.e\ the elliptic curve admitting a complex multiplication of order 3. We will denote by $\varphi:E\ra E$ the automorphism of $E$ given by $(x,y,z)\mapsto (x,y,\zeta z)$, where $\zeta$ is a primitive third root of unity. Let $\EEE$ be the Abelian 3-fold $E\times E\times \ E$ and $\phip$ be the automorphism $\varphi\times \varphi\times \varphi$ acting as $\varphi$ on each factor of $\EEE$. The automorphism $\varphi$ has three fixed points on $E$, which are called  $p_i:=(-1:\zeta^i:0)$, $i=1,2,3$. Hence $\phip$ fixes 27 points on $\EEE$, $p_{i,j,k}:=(p_i;p_j;p_k)$, $i,j,k=1,2,3$. Let $\alpha:\EEE\ra \EEE/\phip$ be the quotient map. The 3-fold $\EEE/\phip$ is singular and its singular locus consists of the 27 points $\alpha(p_{i,j,k})$. Let $\beta:\widetilde{\EEE}\ra \EEE$ be the blow up of $\EEE$ in the 27 points fixed by $\phip$. The exceptional locus consists of 27 disjoint copies of $\mathbb{P}^2$, and the exceptional divisor over the point $p_{i,j,k}$ will be denoted by $\exijk$. The automorphism $\phip$ of $\EEE$ induces the automorphism $\widetilde{\phip}$ on $\widetilde{\EEE}$. Let $\mathcal{Z}:=\widetilde{\EEE}/\widetilde{\phip}$ and $\pi:\widetilde{\EEE}\ra\widetilde{\EEE}/\widetilde{\phip}$ be the quotient map. The  following diagram commutes:
\begin{eqnarray*}
\begin{array}{rcccl}
\phip\circlearrowright&\EEE&\stackrel{\beta}{\longleftarrow}&\widetilde{\EEE}&\circlearrowleft\widetilde{\phip}\\
&\alpha\downarrow&&\downarrow\pi\\
&\EEE/\phip&\stackrel{\gamma}{\longleftarrow}&\mathcal{Z}
\end{array}
\end{eqnarray*}
where $\gamma$ is the contraction of the divisors $\pi(\exijk)$ to the singular points $\alpha(p_{i,j,k})$ of $\EEE/\phip$.\\
The 3-fold $\mathcal{Z}$ is smooth (indeed the fixed locus of $\widetilde{\phip}$ on $\widetilde{\EEE}$ is of codimension 1) and is a Calabi--Yau 3-fold.
\subsection{The cohomology of $\EEE$}\label{section: cohomology of EEE}
The 3-fold $\EEE$ is an Abelian variety. Its canonical bundle is trivial and $$H^{p,q}(\EEE)=
\bigoplus_{\begin{array}{l}a_1+a_2+a_3=p,\\b_1+b_2+b_3=q\end{array}} (H^{a_1,b_1}(E)\otimes H^{a_2,b_2}(E)\otimes H^{a_3,b_3}(E)).$$ Hence the Hodge diamond of $\EEE$ is 
$$\begin{array}{rrrrrrr}
&&&1&\\
&&3&&3\\
&3&&9&&3\\
1&&9&&9&&1\end{array}$$
Let $z_i$ be the complex local coordinate of the $i$-th copy of $E$ in $\EEE$. Then $H^{1,0}(\EEE)=\langle dz_1, dz_2, dz_3\rangle$, $H^{2,0}=\langle dz_1\wedge dz_2, dz_1\wedge dz_3, dz_2\wedge dz_3\rangle$ and $H^{3,0}(\EEE)=\langle dz_1\wedge dz_2\wedge dz_3\rangle$.\\
The Picard group is generated by:
\begin{itemize}\item 3 classes $\Phi_i$, $i=1,2,3$, which are the classes of the fiber of the projection $\rho_i:\EEE\ra E$ on the $i$-th factor, e.g. $\Phi_1=\overline{q}\times E\times E$ for a general point $\overline{q}\in E$; \item 3 classes $\Delta_i$, $i=1,2,3$, which are the product of the $i$-th factor of $\EEE$ by the diagonal on the other two factor, e.g. $\Delta_1=E\times \Delta=\{E\times q\times q| q\in E\}$;\item 3 classes $\Gamma_i$, $i=1,2,3$, which are the product of the $i$-th factor of $\EEE$ by the graph on the other two factors, i.e\ $\Gamma_1=E\times \Gamma=\{E\times q\times \varphi(q)| q\in E\},$ $\Gamma_2=\{\varphi(q)\times E\times q| q\in E\}$, $\Gamma_3=\{q\times \varphi(q)\times E| q\in E\}.$\end{itemize}
By the definition of the divisor $\Phi_i$ it is clear that $\Phi_i=\rho_i^*(\overline{q})$, where $\overline{q}$ is a general point on $E$.\\
A similar description can be given for the divisors $\Delta_i$ and $\Gamma_i$: indeed let $\rho_i:\EEE\ra E$, $\tau_i:\EEE\ra E$ and $\eta_i:\EEE\ra E$ be the maps defined below, then: 
\begin{align}\label{eq: maps asociated to div on EEE} \begin{array}{ll}
\Phi_i=\rho_i^*(P),&\rho_i:(q_1,q_2,q_3)\mapsto q_i\\
\Delta_i=\tau_i^*(P),&\tau_i:(q_1,q_2,q_3)\mapsto q_j-q_k,\ \ \{i,j,k\}=\{1,2,3\}\\
\Gamma_i=\eta_i^*(P),& \eta_i:(q_1,q_2,q_3)\mapsto (\varphi(q_{i+1})- q_{i+2}),\ \ \{i,i+1,i+2\}=\{1,2,3\},\ \ i,i+1,i+2\in\ZZZ.\\\end{array}
\end{align}
We recall that $E\simeq \mathbb{R}^2/\Lambda\simeq \mathbb{C}/\Lambda$, where $\Lambda$ is the lattice generated by $1$ and $\zeta$. Let $(x_{2j-1},x_{2j})$ be the real coordinates of $\mathbb{R}^2$ relative to the $j$-th copy of $E$ and let the isomorphism $\mathbb{R}^2\ra \mathbb{C}$ be $(x_{2j-1},x_{2j})\ra x_{2j-i}+\zeta x_{2j}$, $j=1,2,3$. Each divisor $D$ on $\EEE$ is a linear combination of surfaces on the 3-fold and defines a 2-form, $c_1(D)$, in $H^2(\EEE,\Z)$. The classes of the nine divisors defined above can be found by pulling back the class of a point in $H^2(E,\Z)$. The form associated to a point $q=y_1+\zeta y_2$ on $E$ is $dy_1\wedge dy_2$. We will denote with the same name both the divisor and the corresponding form. So: 
\begin{eqnarray*}\begin{array}{ll}
\Phi_1=\rho_1^*(dy_1\wedge dy_2)=dx_1\wedge dx_2,& \Phi_2=dx_3\wedge dx_4,\\\Phi_3=dx_5\wedge dx_6,\\
\Delta_1=\tau_1^*(dy_1\wedge dy_2)=d(x_3-x_5)\wedge d(x_4-x_6),&\Delta_2=d(x_1-x_5)\wedge d(x_2-x_6),\\ \Delta_3=d(x_1-x_3)\wedge d(x_2-x_4),\\
\Gamma_1=\eta_1^*(dy_1\wedge dy_2)=d(x_5+x_4)\wedge d(x_6-x_3+x_4),&\Gamma_2=d(x_1+x_6)\wedge d(x_2-x_5+x_6),\\ \Gamma_3=d(x_3+x_2)\wedge d(x_4-x_1+x_2).
\end{array} \end{eqnarray*}
In the last two lines we used: $\varphi(q)=\zeta q$, hence $\varphi(x_{2i-1}+\zeta x_{2i})=\zeta x_{2i-1}+(-\zeta-1) x_{2i}=-x_{2i}+\zeta(x_{2i-1}-x_{2i})$.\\
Let us now consider the space $H^{2,2}(\EEE)$. We recall that $H^{2,2}(\EEE)$ is the dual of $H^{1,1}(\EEE)$ and in particular $H^{2,2}(\EEE)\cap H^{4}(\EEE,\Z)$ is generated by nine 4-forms, which are identified (via Poincar\'e duality) with nine 1-cycles. A $\Q$-basis of $H^{2,2}(\EEE)\cap H^{4}(\EEE,\Z)$ generated by classes of curves on $\EEE$, which are the pull back of the class of a general point $Q\in E\times E$ along certain maps $\EEE\rightarrow E\times E$, is:
$$\begin{array}{ll}
\phi_i=\rho_{j,k}^*(Q),&\rho_{j,k}:(q_1,q_2,q_3)\mapsto(q_j,q_k),\ \ i=1,2,3,\ \ \{i,j,k\}=\{1,2,3\}\\
\delta_i=\tau_{j,k}^*(Q),&\tau_{j,k}:(q_1,q_2,q_3)\mapsto (q_i,q_j-q_k),\ \ i=1,2,3,\ \ \{i,j,k\}=\{1,2,3\}\\
\gamma_i=\eta_{j,k}^*(Q)& \eta_{j,k}:(q_1,q_2,q_3)\mapsto (q_i,\varphi(q_{j})- q_{k}),\ \ i=1,2,3,\ \ j=i+1\in\ZZZ, k=i+2\in\ZZZ.\end{array}
$$
One can directly check the following intersection products: 
\begin{align}\label{eq: curves int two divisors EEE}\begin{array}{ccc}\phi_i=\Phi_j\Phi_k,\  \{i,j,k\}=\{1,2,3\},\ \ i=1,2,3,&
\delta_i=\Phi_i\Delta_i,\ \ i=1,2,3,& \gamma_i=\Phi_i\Gamma_i,\ \ i=1,2,3.
\end{array}\end{align}
As $\phi_i=\Phi_j\Phi_k$, its class in $H^4(\EEE,\Z)$ is the wedge product of the 2-forms associated to $\Phi_j$ and $\Phi_k$. 
The intersection between a divisor in $Pic(\EEE)$ and a curve in $H^{2,2}(\EEE)$ is the wedge product of a 4-form and a 2-form, hence it is an element in $H^6(\EEE,\mathbb{Q})\simeq \mathbb{Q}$, where the isomorphism is given by $dx_1\wedge dx_2\wedge dx_3\wedge dx_4\wedge dx_5\wedge dx_6\mapsto 1$. 
From this one finds the intersection numbers between the divisors generating $Pic(\EEE)$ and the curves generating $H^{2,2}(\EEE)$:
\begin{eqnarray}\label{eq: int curves divisors EEE}
\begin{array}{|c|c|c|c|c|c|c|c|c|c|}
\hline
& \Phi_1 & \Phi_2 & \Phi_3 & \Delta_1 & \Delta_2 & \Delta_3 & \Gamma_1 & \Gamma_2 & \Gamma_3 \\
\hline
\phi_1 & 1 & 0 & 0 & 0 & 1 & 1 & 0 & 1 & 1 \\
\hline
\phi_2 & 0 & 1 & 0 & 1 & 0 & 1  & 1 & 0 & 1  \\
\hline
\phi_3 & 0 & 0 & 1 & 1  & 1 & 0 &  1  & 1 & 0  \\
\hline
\delta_1 &  0 & 1 & 1 & 0 & 1 & 1 & 3 & 1 & 1 \\
\hline
\delta_2 &  1 & 0 & 1 & 1 & 0 & 1 & 1 & 3 & 1 \\
\hline
\delta_3 &  1 & 1 & 0 & 1 & 1 & 0 & 1 & 1 & 3 \\
\hline
\gamma_1 &  0 & 1 & 1 & 3 & 1 & 1 & 0 & 1 & 1 \\ 
\hline
\gamma_2& 1 & 0 & 1 & 1 & 3 & 1 & 1 & 0 & 1 \\
\hline
\gamma_3 & 1 & 1 & 0 & 1 & 1 & 3 & 1 & 1 & 0\\
\hline
\end{array}\end{eqnarray}
Now \eqref{eq: curves int two divisors EEE} and \eqref{eq: int curves divisors EEE} together give the cubic self--intersection form on $Pic(\EEE)$:
\begin{align}\label{eq: trilin on EEE}\begin{array}{l}(\sum_{i=1}^3 a_i\Phi_i+b_i\Delta_i+c_i\Gamma_i)^3= 
6 \cdot \Big ( a_1 a_2 a_3 + 
\sum_{i < j }a_i a_j (b_i  + b_j+ c_i  + c_j) + \\ 
\left( \sum_{i < j } b_i b_j \right) \left( a_1 + a_2 + a_3 + 3 \left( c_1 + c_2 + c_3 \right) \right)+ \
\left(  \sum_{i < j } c_i c_j \right) \left( a_1 + a_2 + a_3 + 3  \left( b_1 + b_2 + b_3 \right) \right)+ \\
(a_1 + a_2 + a_3) (b_1 + b_2 + b_3) (c_1 + c_2 + c_3) + 2 \sum_{i=1}^3 a_i b_i c_i  - \sum_{i \neq j} a_i b_j c_j \Big) \, .
\end{array}\end{align}
From the cubic self-intersection form one deduces the trilinear form on $Pic(\EEE)$.

\subsection{The cohomology of $\widetilde{\EEE}$}\label{section: the cohomology of widetildeEEE}
The 3-fold $\widetilde{\EEE}$ is obtained by blowing up the 27 points $p_{i,j,k}:=(p_i;p_j;p_k)$, $i,j,k=1,2,3$, on $\EEE$ and $\exijk$ are the exceptional divisors of this blow up. Hence there is an isomorphism $\widetilde{\EEE}-\bigcup\exijk\simeq \EEE-\bigcup p_{i,j,k}$. The numbers $h^{i,j}$ with $i$ or $j$ equal to zero are birational invariants, hence $h^{i,j}(\widetilde{\EEE})=h^{i,j}(\EEE)$ if $i$ or $j$ are zero.\\
Let $X$ be a projective manifold, $S$ be a codimension $r$ submanifold of $X$ and $\widetilde{X}$ be a blow up of $X$ in $S$. Then:
$H^k(\widetilde{X},\Z)=H^k(X,\Z)\oplus \bigoplus_{i=0}^{r-2}H^{k-2i-2}(S,\Z)$ (\cite[Th\'eor\`eme 7.31]{voisin}). Applying this result to $\widetilde{\EEE}$, the blow up of $\EEE$ in 27 points, we obtain the Hodge diamond of $\widetilde{\EEE}$: 
$$\begin{array}{rrrrrrr}
&&&1&\\
&&3&&3\\
&3&&36&&3\\
1&&9&&9&&1\end{array}$$
Blowing up 27 points, we introduced 27 exceptional divisors $\exijk$, hence $h^{1,1}(\widetilde{\EEE})=h^{1,1}(\EEE)+27$ and $H^{1,1}(\widetilde{\EEE})$ is generated by the 36 classes: 
$$\widetilde{F_h}:=\beta^*(\Phi_h), \ \ \widetilde{D_h}:=\beta^*(\Delta_h),\ \ \widetilde{G_h}:=\beta^*(\Gamma_h),\ \ h=1,2,3,\mbox{ and by the classes }[\exijk],\ \ i,j,k=1,2,3.$$
The divisors $\widetilde{F_i}$, $\widetilde{D_i}$, $\widetilde{G_i}$ are the classes of the strict transforms of $\Phi_i$, $\Delta_i$, $\Gamma_i$, indeed the $\Phi_i$ do not pass through the points $p_{i,j,k}$ and $\Delta_i$, $\Gamma_i$ are cohomologically equivalent to classes which do not pass through $p_{i,j,k}$, for example $\Delta_1$ is in the same class of $\{E\times q\times (q+q')| q\in E\}$ in $H^2(\EEE,\Z)$ for any $q'\in E$.\\
The intersection form on $H^{1,1}(\widetilde{\EEE})=Pic(\widetilde{\EEE})$  is induced by the one on $\EEE$. More precisely:\begin{itemize}
\item
let $L$ be a divisor in $Pic(\EEE)$, then $\beta^*(L)\exijk=0$, indeed all the divisors in $Pic(\EEE)$ are linear combinations of $\Phi_i$, $\Delta_i$ and $\Gamma_i$ and all these divisors are equivalent to divisors which do not pass through the points $p_{ijk}$ and thus their strict transforms do not intersect the exceptional divisors;\item similarly,
for $L_1,\ L_2,\ L_3\in Pic(\EEE)$ we have $\beta^*(L_1)\beta^*(L_2)\beta^*(L_3)=L_1L_2L_3$; \item
$\exijk\widetilde{B_{h,l,m}}=0$ if $(i,j,k)\neq (h,l,m)$, because they are exceptional divisors over distinct points;\item
$\exijk^3=1$ (see \eqref{eq: cube B}).\end{itemize}
Dually the space $H^{2,2}(\widetilde{\EEE})$ is generated by 36 classes. We give a $\Q$-basis of $H^{2,2}(\widetilde{\EEE})\cap H^4(\widetilde{\EEE},\Z)$ made up of curves. Nine of them are the pull-back via $\beta$ of the classes of the curves generating $H^{2,2}(\EEE)$ ($\widetilde{f_i}=\beta^*(\phi_i)$, $\widetilde{d_i}=\beta^*(\delta_i)$, $\widetilde{g_i}=\beta^*(\gamma_i)$) and the other 27 are the classes of the lines $\widetilde{l_{i,j,k}}$ which generate the Picard group of the exceptional divisors $\exijk$.\\ 
By the adjunction formula, the canonical divisor of $\exijk$ is $$K_{\exijk}=\left(\exijk+K_{\widetilde{\EEE}}\right)\exijk=\left(\exijk+\beta^*\left(K_{\EEE}\right)+2\sum_{i,j,k}\exijk\right)\exijk=3\exijk^2.$$ Since $\exijk\simeq \mathbb{P}^2$, $K_{\exijk}=-3\widetilde{l_{i,j,k}}$, and comparing the two expressions of $K_{\exijk}$ we obtain $\widetilde{l_{i,j,k}}=-\exijk^2$. Moreover 
\begin{equation}\label{eq: cube B}1=(-\widetilde{l_{i,j,k}})^2=\exijk_{|\exijk}\cdot \exijk_{|\exijk}=(\exijk\cdot \exijk)_{|\exijk}=\exijk^3.\end{equation}
The intersection form between the curves generating $H^{2,2}(\widetilde{\EEE})$ and the divisors generating $H^{1,1}(\widetilde{\EEE})$ is induced by the one on $\EEE$: if $c\in H^{2,2}(\EEE)$ and $L\in H^{1,1}(\EEE)$ are chosen among the classes appearing in table \ref{eq: int curves divisors EEE}, then: $c\cdot L=\beta^*(c)\cdot \beta^* (L)$; $\beta^*(c)\exijk=\widetilde{l_{i,j,k}}\beta^*(L)=0$; $\widetilde{l_{i,j,k}}\exijk=-\exijk^3=-1$; $\widetilde{l_{h,m,n}}\exijk=0$ if $(h,m,n)\neq (i,j,k)$.

\subsection{The cohomology of $\mathcal{Z}$}\label{section: the cohomology of Z} 
The map $\widetilde{\phip}$ (induced by $\phip$) fixes the divisors $\exijk$ and is without fixed points on $\widetilde{\EEE}-\bigcup_{i,j,k=1,2,3} \exijk$. So the fixed locus $Fix_{\widetilde{\varphi}}(\widetilde{\EEE})=\bigcup_{i,j,k=1,2,3} \exijk$ has codimension 1 and hence the 3-fold $\mathcal{Z}$, which is the quotient $\widetilde{\EEE}/\widetilde{\phip}$, is smooth. Moreover $H^{p,q}(\mathcal{Z})=H^{p,q}(\widetilde{\EEE})^{\widetilde{\phip}}$. We recall that $H^{i,0}(\widetilde{\EEE})=H^{i,0}(\EEE)$ and that the action of $\widetilde{\phip}$ is $(z_1,z_2,z_3)\mapsto (\zeta z_1,\zeta z_2, \zeta z_3)$ (where $z_i$ are the local complex coordinates of the $i$-th copy of $E$). Now it is clear that $\beta^*(dz_i)$ and $\beta^*(dz_i\wedge dz_j)$, $i\neq j$, $i,j=1,2,3$, are not invariant under the action of $\widetilde{\phip}$, but that $\beta^*(dz_1\wedge dz_2\wedge dz_3)$ is invariant under $\widetilde{\phip}$. We conclude that $H^{1,0}(\mathcal{Z})=H^{0,1}(\mathcal{Z})=H^{2,0}(\mathcal{Z})=H^{0,2}(\mathcal{Z})=0$ and $H^{3,0}(\mathcal{Z})=\C \omega_{\mathcal{Z}}$ with $\pi^*\omega_{\mathcal{Z}}=\beta^*(dz_1\wedge dz_2\wedge dz_3)$.\\
Analogously one can compute $H^{2,1}(\mathcal{Z})=H^{2,1}(\widetilde{\EEE})^{\widetilde{\phip}}$. Since $H^{2,1}(\widetilde{\EEE})$ is generated by $\beta^*(dz_i\wedge dz_j\wedge \overline{d z_k})$ for $\{i,j,k\}=\{1,2,3\}$, which are not invariant under $\widetilde{\phip}$, we obtain $H^{2,1}(\mathcal{Z})=H^{1,2}(\mathcal{Z})=0$.\\
The divisors on $\widetilde{\EEE}$ induce divisors on $\mathcal{Z}$. Since the map $(\widetilde{\phip})^*$ acts as the identity on $Pic(\widetilde{\EEE})$, the map $\pi_*:Pic(\widetilde{\EEE})\otimes\Q\ra Pic(\mathcal{Z})\otimes \Q$ is bijective, and hence, as we will see, the Picard group of $\mathcal{Z}$ is generated by the 36 classes  $\pi_*(\widetilde{F_i})$, $\pi_*(\widetilde{D_i})$, $\pi_*(\widetilde{G_i})$, $\pi_*(\exijk)$, $i,j,k=1,2,3$, at least over $\Q$ and we observe that $Pic(\mathcal{Z})\otimes \Q$ is $H^2(\mathcal{Z},\Q)$.  The divisors $\widetilde{F}_i$, $\widetilde{D_i}$, $\widetilde{G_i}$, $\exijk$ generating $Pic(\widetilde{\EEE})$ correspond to surfaces on $\widetilde{\EEE}$. Let us denote by $\widetilde{L}$ one of them, then we define $L$ to be $$L:=\pi(\widetilde{L})\mbox{ as a set, with the reduced scheme structure.}$$ Thus we get the classes $F_i$, $D_i$, $G_i$, $B_{i,j,k}$ which correspond to surfaces on $\mathcal{Z}$. By construction the quotient map $\pi:\widetilde{\EEE}\ra \mathcal{Z}$ is a 3:1 cover branched over $\pi(\bigcup_{i,j,k=1,2,3} \exijk)=\bigcup_{i,j,k=1,2,3} B_{i,j,k}.$ Hence the map $\pi:\exijk\ra B_{i,j,k}$ is $1:1$. Moreover also $\pi:\widetilde{F_i}\ra F_i$, $\pi:\widetilde{D_i}\ra D_i$, $\pi:\widetilde{G_i}\ra G_i$ are $1:1$. Thus
$$\pi_*(\exijk)=B_{i,j,k},\ \ \pi_*(\widetilde{F_i})=F_i,\ \ \pi_*(\widetilde{D_i})=D_i,\ \ \pi_*(\widetilde{G_i})=G_i.$$
The set $\{F_h, D_h, G_h, B_{i,j,k}\}_{h,i,j,k=1,2,3}$ is  a $\Q$-basis of $Pic(\mathcal{Z})$. However it is known that this $\Q$-basis is not a $\Z$-basis. Indeed the class of the branch locus of an $n:1$ cyclic cover is $n$--divisible in the Picard group (cf.\ \cite[Lemma 17.1, Chapter I]{BPV}), in particular there exists a divisor \begin{eqnarray}\label{formula: L}M\in Pic(\mathcal{Z}) \mbox{ such that }3M\simeq \sum_{i,j,k} B_{i,j,k}=:B\end{eqnarray} 
where $A\simeq B$ if the two cycles $A$ and $B$ have the same cohomology class. Of course $M$ is not a linear combination with integer coefficients of the $B_{i,j,k}$.\\
We recall that $Pic(\mathcal{Z})=H^{1,1}(\mathcal{Z})$ and so the Hodge diamond of $\mathcal{Z}$ is  
$$\begin{array}{rrrrrrr}
&&&1&\\
&&0&&0\\
&0&&36&&0\\
1&&0&&0&&1\end{array}$$
The intersection form on $Pic(\mathcal{Z})$ is induced by the one on $Pic(\widetilde{\EEE})$, but one has to recall that the map $\pi:\widetilde{\EEE}\ra \mathcal{Z}$ is a $3:1$ map  away from the ramification locus, where it is a bijection. 
The map $\pi^*:H^*(\mathcal{Z},\C)\ra H^*(\widetilde{\EEE},\C)$ is a homomorphism of rings and hence for each $D,D',D''\in Pic(\mathcal{Z})\subset H^*(\mathcal{Z})$: \begin{equation}\label{eq: int 1}\pi^*(D)\cdot\pi^*(D')=\pi^*(D\cdot D'),\ \ \ \pi^*(D)\cdot\pi^*(D')\cdot\pi^*(D'')=\pi^*(D\cdot D'\cdot D'')=3(D\cdot D'\cdot D''), \end{equation}
where the last equality depends on the degree of $\pi$ (cf.\ \cite[Pag. 9]{De}).\\
So to compute the intersection form on $Pic(\mathcal{Z})$, it suffices to divide the intersection form on $\pi^*(Pic(\mathcal{Z}))\subset Pic(\widetilde{\EEE})$  by 3, here we sketch this computation:
\begin{itemize} \item $\pi^*(B_{i,j,k})=3\exijk$, since $B_{i,j,k}$ are in the ramification locus; \item $\pi^*(F_i)=\widetilde{F_i}+\widetilde{\phip}^*(\widetilde{F_i})+(\widetilde{\phip}^2)^*(\widetilde{F_i})\simeq 3\widetilde{F_i}$, in fact $\phip^*(F_i)$ and $F_i$ have the same cohomology class on $\EEE$ (and hence $\widetilde{\phip}^*(\widetilde{F_i})$ and $\widetilde{F_i}$ have the same cohomology class on $\widetilde{\EEE}$);\item similarly $\pi^*(D_i)\simeq 3\widetilde{D_i}$, $\pi^*(G_i)\simeq 3\widetilde{G_i}$.\end{itemize} Together with the description of the map $\pi_*$ this implies that for every divisor $\widetilde{L}\in Pic(\widetilde{\EEE})$, $\pi^*(L)=\pi^*(\pi_*(\widetilde{L}))\simeq 3\widetilde{L}$.\\  
By \eqref{eq: int 1}, we have $(3\widetilde{L})(3\widetilde{L'})(3\widetilde{L''})=\pi^*(L)\cdot\pi^*(L')\cdot\pi^*(L'')=3(L\cdot L'\cdot L'')$ and thus $$LL'L''=9\widetilde{L}\widetilde{L'}\widetilde{L''}.$$
Hence we obtain that the trilinear form on $\pi_*(Pic(\widetilde{\EEE}))\subset Pic(\mathcal{Z})$ is the trilinear form of $Pic(\widetilde{\EEE})$ multiplied by 9. Since the divisors in $\pi_*(Pic(\widetilde{\EEE}))$ define a $\Q$-basis for $Pic(\mathcal{Z})$, this determines the trilinear form on $Pic(\mathcal{Z})$ completely. To recap, we proved that each divisor $L\in Pic(\mathcal{Z})$ can be written as $L=L_E+L_B$, where $L_E=\sum_{i=1}^3 \left(a_i F_i+b_i D_i+c_iG_i\right)$, $L_B=\sum_{i,j,k=1}^3\alpha_{i,j,k}B_{i,j,k}$ and its cubic self-intersection is  \begin{equation}\label{formula: cubic form}L^3=L_E^3+L_B^3=9\left(\left(\sum_{i=1}^3 \left(a_i \Phi_i+b_i \Delta_i+c_i\Gamma_i\right)\right)^3+\sum_{i,j,k=1}^3\alpha^3_{i,j,k}\right).\end{equation}
We found a $\Q$-basis of $H^{2,2}(\widetilde{\EEE})$ and this induces, via $\pi_*$, a $\Q$-basis of $H^{2,2}(\mathcal{Z})$ (in analogy to what we did for $Pic(\mathcal{Z})$).
Hence a $\Q$-basis for $H^{2,2}(\mathcal{Z})$ consists of the curves $f_i$($=\pi_*(\beta^*(\phi_i))$), $d_i$($=\pi_*(\beta^*(\delta_i))$), $g_i$ ($=\pi_*(\beta^*(\gamma_i))$), $l_{i,j,k}$($=\pi_*(\widetilde{l_{i,j,k}}))$.
The intersection number $aL$ between $a=:\pi_*(\widetilde{a})\in H^{2,2}(\mathcal{Z})$ and $L$, a divisor of the chosen $\Q$-basis of $Pic(\mathcal{Z})$, can be computed by the projection formula (cf.\ \cite[pag. 9]{De}): \begin{equation}\label{eq:H22H11 in Z}aL=\pi_*(\widetilde{a})L=\widetilde{a}\pi^*L=3\widetilde{a}\widetilde{L}\end{equation}
For example, choosing $a=l_{i,j,k}$ and $L=B_{i,j,k}$, we have $l_{i,j,k}B_{i,j,k}=-3$.\\
We will need the expression of certain curves and classes in $H^{2,2}(\mathcal{Z})$ as linear combinations of the classes generating $H^{2,2}(\mathcal{Z})$, so here we compute some of them as examples. 

\begin{ex}\label{rem: the class M2}{\rm {\bf The class $M^2$.} The space $H^{2,2}(\mathcal{Z})$ contains all the classes obtained as intersection of two divisors on $\mathcal{Z}$. In particular the class $M^2=\left(\frac{1}{3}\sum_{i,j,k}B_{i,j,k}\right)^2$ can be written as linear combination of $f_i$, $d_i$, $g_i$, $l_{i,j,k}$  with coefficients in $\Q$, i.e\ $\frac{1}{9}\sum_{i,j,k}\left(B_{i,j,k}^2\right)=(\sum_{h=1}^3(\lambda_hf_h+\mu_h d_h+\nu_h g_h)+\sum_{i,j,k}\alpha_{i,j,k}l_{i,j,k})$. To find the coefficients of this $\Q$-linear combination it suffices to compute the intersection of the divisors $F_i$, $D_i$, $G_i$, $B_{i,j,k}$ with $M^2$. The only non trivial intersections of $M^2$ with these divisors are $M^2B_{i,j,k}=1$. We know that $B_{i,j,k}l_{i,j,k}=-3$, $B_{i,j,k}l_{a,b,c}=0$, if $(a,b,c)\neq (i,j,k)$, $B_{i,j,k}f_h=B_{i,j,k}d_h=B_{i,j,k}g_h=0$ (cf.\ \eqref{eq:H22H11 in Z}). This implies that $1=M^2B_{i,j,k}=-3\alpha_{i,j,k}$ and hence $\alpha_{i,j,k}=-1/3$. Using \eqref{eq: int curves divisors EEE} and \eqref{eq:H22H11 in Z}, the intersections of $M^2$ with $F_h$, $D_h$ and $G_h$ give $\lambda_h=\mu_h=\nu_h=0$.}\end{ex}

\begin{ex}\label{ex: curves Cijk and Aijk}{\rm {\bf The curves $C^k_{i,j}\subset \EEE$ and $A^k_{i,j}\subset\mathcal{Z}$.} Let us consider the curve $C^1_{i,j}=E\times \{p_i\}\times \{p_j\}$ where $i,j=1,2,3$ and $p_i$ are the points fixed by $\varphi$ on $E$. This curve passes through 3 fixed points, $p_{a,i,j}\in \EEE$, $a=1,2,3$. The curve $C^1_{i,j}\subset \EEE$ has the same cohomology class as $E\times \{q\}\times \{r\}$ for two general points $q,r\in E$. In particular the class of the curve $C^1_{i,j}$ is the class $\phi_1$ for each $i,j$. Let $\widetilde{C^1_{i,j}}:=\overline{\beta^{-1}(C^1_{i,j})-\cup_{a=1,2,3}\{p_{a,i,j}\}}$, it is the strict transform of $C^1_{i,j}$. The curve $\widetilde{C^1_{i,j}}$ intersects the exceptional divisors $\widetilde{B_{a,b,c}}$ in one point if and only if $(b,c)=(i,j)$. Hence $\widetilde{C^1_{i,j}}=\beta^*(\phi_1)-\sum_{a=1}^3\widetilde{l_{a,i,j}}$. So $\pi_*(\widetilde{C^1_{i,j}}))=\pi_*(\beta^*(\phi_1)-\sum_{a=1}^3\widetilde{l_{a,i,j}})=f_1-\sum_{a=1}^3l_{a,i,j}$. Let us consider $A^1_{i,j}=\pi(\widetilde{C^1_{i,j}}))$ as set with the reduced scheme structure. The map $\pi:\widetilde{C^1_{j,k}}\ra A^1_{i,j}$ is $3:1$, hence $A^{1}_{i,j}=\frac{1}{3}(\pi_*(\beta^*(C^1_{i,j})))$. More generally, define $C^2_{i,j}=\{p_i\}\times E\times \{p_j\}$, $C^3_{i,j}=\{p_i\}\times \{p_j\}\times E$, $A^h_{i,j}=\pi(\widetilde{C^h_{i,j}})$ as a set with the reduced scheme structure, then:
\begin{equation}\label{eq_ the curves Ahij}A^1_{i,j}=\frac{1}{3}\left(f_1-\sum_{a=1}^3l_{a,i,j}\right),\ \ A^2_{i,j}=\frac{1}{3}\left(f_2-\sum_{a=1}^3l_{i,a,j}\right),\ A^3_{i,j}=\frac{1}{3}\left(f_3-\sum_{a=1}^3l_{i,j,a}\right).\end{equation}}\end{ex}

\subsection{The Chern classes and the Riemann--Roch theorem on $\mathcal{Z}$}
The $i$-th Chern class of a variety is the  $i$-th Chern class of its tangent bundle. For a smooth projective variety $X$,  $c_i(X)\in H^{2i}(X,\Z)$ and, by convention, $c_0(X)=1$.\\ 
If $X$ is a Calabi--Yau variety, then $c_1(X)=0$, indeed $c_1(\mathcal{T}_X)=c_1(\bigwedge^3\mathcal{T}_X)=c_1(-K_X)=0$.\\ 
The third Chern class of a smooth projective variety of dimension 3 satisfies $\chi(X)=c_3(X)$ (Gauss--Bonnet formula, \cite[Pag. 416]{GH}).\\ 
Here we compute the Chern classes of $\mathcal{Z}$. From the previous considerations it follows immediately that: $$c_0(\mathcal{Z})=1,\ \ \ c_1(\mathcal{Z})=0,\ \ \ c_2(\mathcal{Z})=\sum_{h=1}^3(\lambda_hf_h+\mu_h d_h+\nu_h g_h)+\sum_{i,j,k=1}^3\alpha_{i,j,k}l_{i,j,k},\ \ \ c_3(\mathcal{Z})=\chi(\mathcal{Z})=72,$$
where $\lambda_h$, $\mu_h$, $\nu_h$, $\alpha_{i,j,k}\in \Q$.
It remains to determine the coefficients of the linear combination defining $c_2$. To do this we need the following result: 
\begin{lemma}{\rm (\cite[Lemma 4.4]{Fr})}  If $X$ is a complex 3-fold with trivial canonical bundle and $S$ is a smooth complex surface in $X$, then $c_2(X)[S]=-c_1(S)^2+c_2(S)$.\end{lemma}
We now apply this result to each generator of $Pic(\mathcal{Z})$. As the divisors $B_{i,j,k}$ are isomorphic to $\mathbb{P}^2$, one has $c_1(B_{i,j,k})=3l_{i,j,k}$ and $c_2(B_{i,j,k})=\chi(\mathbb{P}^2)=3$. The divisors $F_i$, $D_i$, $G_i$ are isomorphic to the Abelian surface $E\times E$ (indeed the map $\pi$ is $1:1$ between the Abelian surface $\widetilde{F_i}\simeq E\times E$ and $F_i$ are isomorphic to $E\times E$, similarly $D_i$ and $G_i$). Hence their first Chern class is zero (since their canonical bundle is trivial) and their second Chern class is 0 (since it is equal to their Euler characteristic).\\ Now we compute the coefficients in $c_2(\mathcal{Z})$ as in Example \ref{rem: the class M2}. Indeed using \eqref{eq: int curves divisors EEE} and \eqref{eq:H22H11 in Z} one has:
\begin{eqnarray*}\begin{array}{c}
-6=-c_1(B_{x,y,z})^2+c_2(B_{x,y,z})=c_2(\mathcal{Z})[B_{x,y,z}]=\\ =\left(\sum_{h=1}^3(\lambda_hf_h+\mu_h d_h+\nu_h g_h)+
\sum_{i,j,k}\alpha_{i,j,k}l_{i,j,k}\right)[B_{x,y,z}]=-3\alpha_{x,y,z},\end{array}\end{eqnarray*}
and
\begin{eqnarray*}\begin{array}{c}0=-c_1(F_1)^2+c_2(F_1)=c_2(\mathcal{Z})[F_1]=\\=\left(\sum_{h=1}^3(\lambda_hf_h+\mu_h d_h+\nu_h g_h)+\sum_{i,j,k}\alpha_{i,j,k}l_{i,j,k}\right)[F_1]=3(\lambda_1+\mu_2+\mu_3+\nu_2+\nu_3)\end{array}\end{eqnarray*}
Doing this for all divisors we obtain $\lambda_h=\mu_h=\nu_h=0$, thus:
\begin{equation*}
c_2(\mathcal{Z})=2\sum_{i,j,k=1}^3l_{i,j,k}.\end{equation*}
\begin{rem}{\rm Considering Example \ref{rem: the class M2}, it is immediate to see that the second Chern class $c_2(\mathcal{Z})$ is divisible by 6 in $H^4(\mathcal{Z},\Z)$, indeed $c_2(\mathcal{Z})/6=\sum_{i,j,k=1,2,3}l_{i,j,k}/3=-M^2\in H^4(\mathcal{Z},\Z)$. The divisibility of this class was already obtained in a different and more involved way by Lee and Oguiso, \cite{Oguiso 3}.}\end{rem}
The computation of the second Chern class of $\mathcal{Z}$ allows also to write down explicitly the Riemann-Roch theorem for the divisors on $\mathcal{Z}$. Indeed it is well known (cf.\ \cite{Hartshorne}) that the Riemann-Roch theorem for a 3-fold is:  
$$\chi(\mathcal{L}(D))=\frac{1}{12}D\cdot(D-K_X)(2D-K_X)+\frac{1}{12}c_2\cdot D+1-p_a.$$
In case $X=\mathcal{Z}$ we have $K_{\mathcal{Z}}=0$ and $p_a=1$ (since $\mathcal{Z}$ is a Calabi--Yau variety), so we obtain:
\begin{equation}\label{formula: Riemann Roch Z}\chi(\mathcal{L}(D))=\frac{1}{6}D^3+\frac{1}{6}\sum_{i,j,k=1,2,3}l_{i,j,k}D.{\bf}\end{equation}

\section{More on the trilinear form on $Pic(\EEE)$}\label{sec: more on the trilinear form on Pic(EEE)}
To compute the Yukawa coupling on $\mathcal{Z}$ it is necessary to describe the trilinear form on $Pic(\mathcal{Z})$. We proved in the previous section that the trilinear form of $Pic(\mathcal{Z})$ depends on the trilinear form on $Pic(\EEE)$ (cf.\ \eqref{formula: cubic form}, \eqref{eq: trilinear L cube}). For this reason we now give a different description of the trilinear form on $Pic(\EEE)$: we reduce the computation of this trilinear form to the computation of a determinant of a matrix in $Mat_{3,3}(\Q[\zeta])$ (cf.\ \ref{eq: trilinear and det}). More precisely in this section we give a way to associate to each divisor $L$ on $\EEE$ a matrix $\Omega_L$ in $Mat_{3,3}(\Q[\zeta])$ and we prove the following theorem:
\begin{theorem}\label{theorem: from L to OmegaL} There exists a homomorphism of groups $\mu: Pic(\EEE)\ra \{H\in Mat_{3,3}(\Q[\zeta])\ |\ ^t\overline{H}=-H\}\subset Mat_{3,3}(\Q[\zeta])$ such that, for each divisor $L\in Pic(\EEE)$, $L^3=\frac{1}{12}\sqrt{-3}det (\mu(L))$. \end{theorem} 
We already said (Section \ref{section: cohomology of EEE}) that a divisor $D$ on an Abelian variety $A\simeq \mathbb{R}^n/\Lambda$ corresponds to a 2-form $c_1(D)$ and hence to a skew--symmetric form $E_D$ on the lattice $\Lambda$ taking values in $\Z$.

The elliptic curve $E$ is obtained as $\mathbb{C}/\Z[\zeta]$. Since we are considering the Abelian variety $\EEE$, in this context $\Lambda\simeq \Z[\zeta]^3$, and we are saying that each divisor $D$ in $Pic(\EEE)$ defines a skew-symmetric form $E_D:\Z[\zeta]^3\times \Z[\zeta]^3\ra \Z$.\\
First of all we prove that for each $L\in Pic(\EEE)$ there exists a matrix $\Omega_L$, such that for each $x,y\in \Z[\zeta]^3$, $E_L(x,y)=Tr(^tx\Omega_L \overline{y})$, where $Tr$ is the trace of an element in $\Q[\zeta]$ over $\Q$ defined as $Tr(a+\zeta b)=(a+\zeta b)+(\overline{a+\zeta b})=2a-b$ for $a,b\in \Q$. Since $E_L(x,y)=-E_L(y,x)$, $^T\overline{\Omega_L}=-\Omega_L$.\\
To compute the matrix $\Omega_L$ for each of the nine divisors $L$ which generate the Picard group of $\EEE$, we use the same technique considered in Section \ref{section: cohomology of EEE}, i.e\ we consider divisors which generate $Pic(\EEE)$ as pull-back of divisors on an elliptic curve $E$. Indeed, since the map $c_1:Pic(X)\ra H^2(X,\Z)$ commutes with the pull back, we have that if $L\in Pic(\EEE)$ is $\alpha^*(l)$ for a certain map $\alpha:\EEE\ra E$ and a certain divisor $l\in Pic(E)$, then $E_L(x,y)=E_l(\alpha(x),\alpha(y))$.\\
Let us consider the elliptic curve $E=\C/\Z[\zeta]$ and a general point $P\in E$. Then $c_1(P)$ is the skew-form $E_{P}=\left[\begin{array}{rr}0&1\\-1&0\end{array}\right]$ (the unique, up to a constant, skew-form on $\Lambda$). So $E_{P}(a+\zeta b,c+\zeta d)=ad-bc$. The matrix $\Omega_{P}$ is a $1\times 1$ matrix with entries in $\Q[\zeta]$ (i.e\ $ \Omega_{P}\in \Q[\zeta]$) such that $$Tr((a+\zeta b)\Omega_L(c+\overline{\zeta}d))=(a,b)\left[\begin{array}{rr}0&1\\-1&0\end{array}\right]\left(\begin{array}{l}c\\d\end{array}\right)=ad-bc.$$
This gives $\Omega_{P}=\rho$ (=$-\overline{^t\Omega_{P}}$) where $\rho:=(\zeta-\overline{\zeta})/3$.\\
The matrix $\Omega_L$ for a certain divisor $L\in Pic(\EEE)$ was identified by the property $E_L(v,w)=Tr(^tv\Omega_L\overline w)$, hence to compute it, we consider $\alpha:\EEE\ra E$, $\alpha:(z_1,z_2,z_3)\mapsto \sum_{i=1}^3 a_iz_i$, where $z_i$ are the complex coordinates on the $i$-th copy of $E$. Let $a:=(a_1,a_2,a_3)$ and analogously $v:=(v_1,v_2,v_3)$, $w:=(w_1,w_2,w_3)$. Let $L=\alpha^*(P)$, $\alpha(v)=\sum_ia_iv_i=a^tv$, $\alpha(w)=\sum_ia_iw_i=a^tw$ (where $^tb$ is the transpose of the vector $b$). Then
\begin{equation*}E_L(v,w)=E_{P}(\alpha(v),\alpha(w))=E_{P}(a^tv,a^tw)=Tr(^t(a^tv)\rho\overline{a^tw})=Tr(v^ta\rho\overline{a^tw}).\end{equation*}
which implies that the matrix $\Omega_L$ associated to $L=\alpha^*(P)$ is \begin{equation}\label{eq: OmegaL}\Omega_L:=\rho\left(\begin{array}{ccc}a_1\\a_2\\a_3\end{array}\right)(\overline{a_1},\overline{a_2},\overline{a_3})\end{equation}
Thus to find, for example, $\Omega_{\Gamma_1}$ it suffices to apply \eqref{eq: OmegaL} to the map $\eta_1:(z_1,z_2,z_3)\mapsto \zeta z_2-z_3$:
$$\Omega_{\Gamma_1}=\rho\left(\begin{array}{ccc}0\\\zeta\\-1\end{array}\right)(0,\overline{\zeta},-1)=\rho\left[\begin{array}{rrr}0&0&0\\0&1&-\zeta\\0&-\overline{\zeta}&1\end{array}\right].$$
Similarly one finds $\Omega_{\Phi_i}$, $\Omega_{\Delta_i}$ and $\Omega_{\Gamma_i}$ (the map associated to each of these divisors is given in \eqref{eq: maps asociated to div on EEE}). In this way one finds that if $L=\sum_{i=1}^3(a_i\Phi_i+b_i\Delta_i+c_i\Gamma_i)$, then $\Omega_L=a_i\Omega_{\Phi_i}+b_i\Omega_{\Delta_i}+c_i\Omega_{\Gamma_i}$ is given by
\begin{equation}\label{eq: trilinear and det}\begin{array}{c}\Omega_L=\rho\left[\begin{array}{ccc}a_1+b_2+b_3+c_2+c_3&-b_3-\zeta c_3&-b_2-\overline{\zeta} c_2\\-b_3-\overline{\zeta} c_3&a_2+b_1+b_3+c_1+c_3&-b_1-\zeta c_1\\-b_2-\zeta c_2&-b_1-\overline{\zeta} c_1&a_3+b_1+b_2+c_1+c_2\end{array}\right].\end{array}
\end{equation}
Now an explicit computation shows that for each divisor $L\in Pic(\EEE)$,  the determinant of $\Omega_L$ is, up to a constant, the intersection form computed in \eqref{eq: trilin on EEE},  \begin{equation}\label{eq: trilinear L cube}L^3=\frac{1}{12}\sqrt{-3} det(\sum_{i=1}^3(a_i\Omega_{\Phi_i}+b_i\Omega_{\Delta_i}+c_i\Omega_{\Gamma_i}))\end{equation} and this concludes the proof of Theorem \ref{theorem: from L to OmegaL}.
\begin{rem}{\rm The compatibility between the group structures of $Pic(\EEE)$ and $Mat_{3,3}(\Q[\zeta])$ is due to the properties of the skew symmetric form $E_L$ defined by a divisor $L$ and of the trace $Tr$. Indeed $E_{L\otimes M}(x)=E_L(x)+E_M(x)=Tr(^tx\Omega_L\overline{x})+Tr(^tx\Omega_M\overline{x})=Tr(^tx(\Omega_L+\Omega_M)\overline{x})$, and so to the line bundle $L\otimes M$ we associate the matrix $\Omega_L+\Omega_M$.}\end{rem}

We observe that the Picard group of the singular quotient $\EEE/\phip$ has rank 9 and is induced by the one on $Pic(\EEE)$. The Picard group of $\EEE/\phip$ can be identified with the subgroup of $Pic(\mathcal{Z})$ generated by $F_i$, $D_i$, $G_i$. In Section \ref{section: the cohomology of Z} we proved that the trilinear intersection form on $\langle F_i,D_i,G_i\rangle_{i=1,2,3}\subset Pic(\mathcal{Z})$ is, up to a constant (multiplication by 9), the trilinear form on $Pic(\EEE)$. We deduce that the trilinear form on $Pic(\EEE/\phip)$ is, up to a constant, the determinant of the matrix as in \eqref{eq: trilinear and det}. Since $H^{2,0}(\EEE/\phip)=0$, $Pic(\EEE/\phip)\otimes \C\simeq H^{1,1}(\EEE/\phip)$ and the cup product on $H^{1,1}(\EEE/\phip)$ coincides with the $\C$-linear extension of the trilinear form. So the cup product on $H^{1,1}(\EEE/\phip)$ can be represented as determinant of a matrix in $Mat_{3,3}(\Q[\zeta])$. This is of a certain interest because of its relation with the Yukawa coupling on $H^{1,1}(\EEE/\phip)$, obtained as the sum of the cup product and another addend, involving the Gromov--Witten invariants. The values of $a_i$, $b_i$, $c_i$ where the determinant of the matrix \eqref{eq: trilinear and det} is zero correspond to $(1,1)$ forms where the cup product is zero. The set of such values is described by the cubic $\mathcal{C}_3:=V(det(\sum_{i=1}^3(a_i\Omega_{\Phi_i}+b_i\Omega_{\Delta_i}+c_i\Omega_{\Gamma_i})))$ in $\mathbb{P}^8$ (projective space with coordinates $(a_1:a_2:a_3:b_1:b_2:b_3:c_1:c_2:c_3)$). This cubic is singular where the matrix $\sum_{i=1}^3(a_i\Omega_{\Phi_i}+b_i\Omega_{\Delta_i}+c_i\Omega_{\Gamma_i})$ has rank 1, hence along  the intersection of the nine quadrics in $\mathbb{P}^8$ defined by requiring that the nine $2\times 2$ minors of the matrix are zero. The matrices of rank 1 are of type $\rho (^ta)(\overline{a})$ for a certain vector $a=(a_1,a_2,a_3)$. We already showed that the matrix associated to the divisors $\Phi_i$, $\Delta_i$, $\Gamma_i$ are of this type and hence they correspond to singular points of the cubic. We notice that these divisors define a fibration on the 3-fold.

\begin{rem}{\rm Let $Y$ be a Calabi--Yau 3-fold. In \cite{W}, the cubic hypersurface $W$ in $\mathbb{P}(Pic(Y)\otimes \C)$, consisting of the points representing divisors $L$ with $L^3=0$, is analyzed. Here we are considering the cubic $\mathcal{C}_3$ defined in the same way as $W$, but in the case of the Abelian variety $\EEE$. By the relations between $Pic(\EEE)$ and $Pic(\mathcal{Z})$ given in Section \ref{section: the cohomology of Z}, the cubic $\mathcal{C}_3$ is also related to the cubic $W$ in case $Y=\mathcal{Z}$.}\end{rem}

\section{Projective models of $\mathcal{Z}$}\label{sec: projective models Z}

The aim of this section is to give explicit relations and equations for $\mathcal{Z}$. To do this we describe some (singular) projective models of the 3-fold $\mathcal{Z}$ and more in general maps $f:\mathcal{Z}\ra \mathbb{P}^N$.  Each of these maps is associated to a line bundle $L:=f^*(\mathcal{O}_{\mathbb{P}^N}(1))$, and hence $f$ is given by global sections $s_0,\ldots s_N\in H^0(\mathcal{Z},L)$, i.e\ $f:z\mapsto (s_0(z):\ldots s_N(z)).$\\
Our strategy will be to construct line bundles $L$ (and maps $m_L$ associated to $L$) on $\EEE$ and use these to induce line bundles (and hence maps) on $\mathcal{Z}$.  Let $L$ be a line bundle on $\EEE$ such that $\phip^*L\simeq L$. Then $\phip^*$ acts on $H^0(\EEE,L)$ and hence the space $H^0(\EEE,L)$ is naturally decomposed in three eigenspaces $H^0(\EEE,L)_0$, $H^0(\EEE,L)_1$, $H^0(\EEE,L)_{2}$. By construction the maps $(m_L)_0:\EEE\ra \mathbb{P}(H^0(\EEE,L)_0)$, $(m_L)_1:\EEE\ra \mathbb{P}(H^0(\EEE,L)_1)$, $(m_L)_{2}:\EEE\ra \mathbb{P}(H^0(\EEE,L)_{2})$ identify points on $\EEE$ which are in the same orbit for $\phip$. This implies that these maps (or better the maps induced by these maps on $\widetilde{\EEE}$) are well defined on $\mathcal{Z}$ and thus are associated to line bundles on $\mathcal{Z}$. It is moreover clear that the map $(m_L)_\epsilon$, $\epsilon=0,1,2$, is the composition of $\EEE\ra m_L(\EEE)$ followed by the projection of $m_L(\EEE)$ on the subspace $\mathbb{P}(H^0(\EEE,L))_\epsilon$. First of all we point out the relations between the line bundle and its global sections on $\EEE$ and on $\mathcal{Z}$ and then we focus our attention on a specific case.
\begin{rem}{\rm We said that the space $H^0(\EEE,L)$ is naturally decomposed in eigenspaces by the action of $\phip^*$, and indeed there are three subspaces of $H^0(\EEE,L)$ such that the action of $\phip$ is the same on all the elements in the same subspace and is different on two elements chosen in two different subspaces. However the choice of the eigenvalue of each eigenspace is not canonical, but depends on the lift of $\phip^*$ on $H^0(\EEE,L)$ chosen.}\end{rem}
\begin{lemma}{\rm (\cite[Lemma I.17.2]{BPV})} Let $\pi:X\ra Y$ be an $n$-cyclic covering of $Y$ branched along a smooth divisor $C$ and determined by $\mathcal{O}_Y(L)$, where $L$ is a divisor such that $\mathcal{O}_Y(nL)\simeq \mathcal{O}_Y(C)$. Then $\pi_*(\mathcal{O}_X)=\oplus_{k=0}^{n-1}\mathcal{O}_Y(-kL)$.\end{lemma}
Since $\widetilde{\phip}$ acts as a multiplication by $\zeta$ on the local equation of each ramification divisor $\exijk$, we can apply the previous lemma to $X=\widetilde{\EEE}$, $Y=\mathcal{Z}$, $C=B$, $L=M$ (cf.\ \eqref{formula: L}), obtaining \begin{equation}\label{formula: mathcalOEEE=mathcalOZ}\pi_*(\mathcal{O}_{\widetilde{\EEE}})=\mathcal{O}_{\mathcal{Z}}\oplus \mathcal{O}_{\mathcal{Z}}(-M)\oplus \mathcal{O}_{\mathcal{Z}}(-2M).\end{equation} Indeed $\mathcal{O}_{\mathcal{Z}}$, $\mathcal{O}_{\mathcal{Z}}(-M)$, $\mathcal{O}_{\mathcal{Z}}(-2M)$ correspond to the subbundles of $\pi_*(\mathcal{O}_{\widetilde{\EEE}})$ which are stable with respect to the action of $\widetilde{\varphi_3}$. In particular, $\mathcal{O}_{\mathcal{Z}}$ corresponds to the subbundle of the eigenvalue 1. \\
Let $\widetilde{L}\in Pic(\widetilde{\EEE})$ be such that there exists $L\in Pic(\mathcal{Z})$ satisfying $\widetilde{L}=\pi^*(L)$. Then 
\begin{equation}\label{formula: pi}\pi_*(\widetilde{L})=\pi_*(\pi^*(L)\otimes \mathcal{O}_{\mathcal{Z}})=L\otimes \pi_*\mathcal{O}_{\mathcal{Z}}=L\oplus L(-M)\oplus L(-2M)\end{equation} where the last equality follows from \eqref{formula: mathcalOEEE=mathcalOZ}. This implies that \begin{equation}\label{formula: H0EEE=H0Z}H^0(\widetilde{\EEE},\widetilde{L})=H^0(\mathcal{Z},\pi_*(\widetilde{L}))=H^0(\mathcal{Z},L)\oplus H^0(\mathcal{Z}, L-M)\oplus H^0(\mathcal{Z}, L-2M)\end{equation}
where in the first equality we used, viewing $\widetilde{L}$ as invertible sheaf, $H^0(\mathcal{Z},\pi_*(\widetilde{L}))=(\pi_*(\widetilde{L}))(\mathcal{Z})=\widetilde{L}(\pi^{-1}(\mathcal{Z}))=\widetilde{L}(\widetilde{\EEE})=H^0(\widetilde{\EEE}, \widetilde{L})$ and in the last \eqref{formula: pi}.\\

Now we concentrate on a specific choice of divisors on $\EEE$ and $\mathcal{Z}$: let $\Phi:=\Phi_1+\Phi_2+\Phi_3\in Pic (\EEE)$ and $F:=F_1+F_2+F_3\in Pic(\mathcal{Z})$.
\begin{prop}\label{prop: F, F-M, F-2m} The map $m_{3\Phi}:\EEE\ra \mathbb{P}^{26}$ is an embedding. The automorphism $\phip$ of $\EEE$ extends to an automorphism, called again $\phip$, on $\mathbb{P}^{26}$. Let $(\mathbb{P}^{26})_{\epsilon}$ be the eigenspace for the eigenvalue $\zeta^\epsilon$, $\epsilon=0,1,2$ for $\phip$. The composition of $m_{3\Phi}$ with projection $\mathbb{P}^{26}\ra \mathbb{P}^{26}_0$ (resp. $\mathbb{P}^{26}_{1}$, $\mathbb{P}^{26}_{2}$) is the map defined on $\mathcal{Z}$ which is associated to the divisor $F$ (resp. $F-M$, $F-2M$).\end{prop}
\proof
The diagram: $$\begin{array}{ccl}\EEE&\stackrel{\beta}{\leftarrow} &\widetilde{\EEE}\\&&\downarrow\pi\\&&\mathcal{Z}\end{array}\mbox { induces }\begin{array}{ccl}H^0(\EEE,3\Phi)&\stackrel{\beta^*}{\rightarrow} &H^0(\widetilde{\EEE},3\beta^*(\Phi))\\&&\uparrow\pi^*\\&&H^0(\mathcal{Z},F)\end{array}$$
The map $\beta^*$ is an isomorphism. A section $s\in H^0(\EEE,3\Phi)$ with divisor $D$ which has multiplicity $\alpha_{i,j,k}$ in the point $p_{i,j,k}$ pulls back to a section $\beta^*s$ with divisor $\beta^*D=\widetilde{D}+\sum_{i,j,k}\alpha_{i,j,k}\exijk$, where $\widetilde{D}$ is the strict transform of $D$. Since $3\beta^*(\Phi)=\pi^*(F)$, using \eqref{formula: pi} and \eqref{formula: H0EEE=H0Z}, we have $\pi_*(3\beta^*(\Phi))=F\otimes \pi_*\mathcal{O}_{\mathcal{Z}}$ and $$H^0(\widetilde{\EEE},3\beta^*(\Phi))=H^0(\mathcal{Z},F\otimes \pi_*\mathcal{O}_{\mathcal{Z}})=H^0(\mathcal{Z}, F)\oplus H^0(\mathcal{Z}, F(-M))\oplus H^0(\mathcal{Z}, F(-2M)).$$ Thus \begin{equation}\label{eq: eigenspace}H^0(\EEE,3\Phi)\stackrel{\beta^*}{\simeq}H^0(\widetilde{\EEE},3\beta^*\Phi)\simeq H^0(\mathcal{Z}, F)\oplus H^0(\mathcal{Z}, F-M)\oplus H^0(\mathcal{Z}, F-2M)\end{equation} and the last decomposition is a decomposition in eigenspaces of $H^0(\EEE,3\Phi)$. So $H^0(\mathcal{Z},F-aM)\subset H^0(\EEE,3\Phi)$ corresponds to the space of the sections of $3\Phi$ on $\EEE$ with zeros with multiplicity at least $a$ in the points $p_{i,j,k}$ and which are in the same eigenspace for $\phip$.
The map associated to $3\Phi$ is very explicit: Every elliptic curve is embedded in $\mathbb{P}^2$ as a cubic, by the linear system associated to the divisor $3P$. In particular the curve $E$ has  the curve $x^3+y^3+z^3=0$ as image in $\mathbb{P}^2_{x,y,z}$. So we can embed $\EEE$ in $\mathbb{P}^2_{x_1,y_1,z_1}\times \mathbb{P}^2_{x_2,y_2,z_2}\times \mathbb{P}^2_{x_3,y_3,z_3}$ (embedding each factor of $\EEE$ in the corrispondent copy of $\mathbb{P}^2$). Now it is well known that there exists an embedding of $\mathbb{P}^2\times \mathbb{P}^2\times \mathbb{P}^2$ in $\mathbb{P}^{26}$ given by the Segre map $$s:((x_1:y_1:z_1),(x_2:y_2:z_2),(x_3:y_3:z_3))\ra (x_1x_2x_3:x_1x_2y_3:x_1x_2z_3:x_1y_2x_3:\ldots: z_1z_2z_3).$$ Hence there is an embedding of $\EEE$ in $\mathbb{P}^{26}$ which is the restriction of $s$ to $\EEE$. By construction this map is associated to the very ample divisor  $3\Phi$ on $\EEE$. This map extends to a map defined on $\widetilde{\EEE}$ which contracts the exceptional divisors $\exijk$ (which are in fact orthogonal to the divisor $\beta^*(3\Phi)$ defining the map).\\
The action of the automorphism $\phip$ on $\EEE$ is given by $\varphi:((x_1:y_1:z_1),(x_2:y_2:z_2),(x_3:y_3:z_3))\ra ((x_1:y_1:\zeta z_1),(x_2:y_2:\zeta z_2),(x_3:y_3:\zeta z_3))$ and this automorphism extends to an automorphism on $\mathbb{P}^{26}$. The eigenspaces with eigenvalue $1$, $\zeta$, $\zeta^2$ for $\phip$ on $\mathbb{P}^{26}$ are 
$$\begin{array}{c}(x_1x_2x_3:x_1x_2y_3:x_1y_2x_3:x_1y_2y_3:y_1x_2x_3:y_1x_2y_3:y_1y_2x_3:y_1y_2y_3:z_1z_2z_3)\\
(z_1 x_2 x_3: z_1 x_2 y_3 : z_1 y_2 x_3 : z_1 y_2 y_3 : x_1 z_2 x_3 : x_1 z_2  y_3 : y_1 z_2 x_3 : y_1 z_2 y_3 : x_1 x_2 z_3 : x_1 y_2 z_3 : y_1 x_2 z_3 : y_1 y_2 z_3 )\\
(x_1z_2z_3:y_1z_2z_3:z_1x_2z_3:z_1y_2z_3:z_1z_2x_3:z_1z_2y_3)\end{array}$$
respectively. We observe that the first eigenspace is defined by sections of $3\Phi$ which are not necessarily zero in the points $p_{i,j,k}$ (for example the monomial $x_1x_2x_3$ is not zero in the points $p_{i,j,k}$), the second by sections passing through $p_{i,j,k}$ with multiplicity 1 and the third by sections passing through the points $p_{i,j,k}$ with multiplicity 2. Hence the first eigenspace is identified (under the isomorphisms \eqref{eq: eigenspace}) with $H^0(\mathcal{Z},F)$, the second with $H^0(\mathcal{Z},F-M)$ and the third with $H^0(\mathcal{Z},F-2M)$.\endproof
\begin{rem}\label{rem: F-aM and eigenspaces}{\rm  From this description of $H^0(\mathcal{Z},F-kM)$ we get $$\dim(H^0(\mathcal{Z},F))=9,\ \ \dim(H^0(\mathcal{Z},F-M))=12,\ \ \dim(H^0(\mathcal{Z},F-2M))=6.$$
If we apply the Riemann--Roch theorem (cf.\ \eqref{formula: Riemann Roch Z}) to the divisor $F$, $F-M$, $F-2M$, we find $\chi(F)=9$, $\chi(F-M)=12$, $\chi(F-2M)=6$. This in particular implies that for a divisor $L$ among $F$, $F-M$, $F-2M$, $h^2(\mathcal{Z},L)-h^1(\mathcal{Z},L)=0$, indeed by Serre duality we have $h^3(\mathcal{Z},L)=0$. For the divisors $F$ and $F-M$ this is a trivial consequence of the fact that they are big and nef, as we will see in Propositions \ref{prop: m_0} and \ref{prop: m_1},  and of the Kawamata--Viehweg vanishing theorem.}\end{rem}

\begin{rem}\label{rem: hF-kM}{\rm Analogously we can consider the sections of the line bundles $hF-kM$, $h,k\geq 0$, over $\mathcal{Z}$. These correspond (as showed for $F-kM$) to sections of $3h\Phi$ over $\EEE$ which vanish at least of degree $k$ in the points $p_{i,j,k}$. We denote by $N_{h,k}$ the space of such a sections. In case $k=0,1,2$ this gives a decomposition in eigenspaces of $H^0(\EEE,h\Phi)$ relative to the eigenvalue $\zeta^k$.\\
Let $k=0,1,2$. We denote by $\left(\mathit{Sym}^h(E)\right)_k :=  \left(\mathit{Sym}^h<x,y,z>\right)_k $ the monomials of degree $h$ in the coordinates of $E$ which belong to the eigenspace of the eigenvalue $\zeta^k$. Now $(\mathit{Sym}^h(E))_k$ is generated by the monomials of the form $x^\alpha y^\beta z^\gamma$ such that $\alpha + \beta + \gamma =h$ and $\gamma\equiv k\mod 3$. Since $z^3=-x^3-y^3$ on $E$, we can assume that $\gamma=k$. Thus the eigenspaces have the following dimensions: $dim (\mathit{Sym}^h(E) )_k = h+1$, $h$, $h-1$ for $k=0,1,2$ respectively. The sections of $3h\Phi$ on $\EEE$ are given by $\mathit{Sym}^h(E) \times \mathit{Sym}^h(E) \times \mathit{Sym}^h(E)$ and hence $N_{h,k}=\mathit{Sym}^h(E)_a \times \mathit{Sym}^h(E)_b\times \mathit{Sym}^h(E)_c$ with $a,b,c=0,1,2$ and $a+b+c\equiv k\mod 3$. After direct computation we obtain the following dimensions
$$
\begin{array}{rcl}
dim (N_{h,k})= \left\{ 
\begin{array}{ll}
9 h^3 & k=0\\
9 h^3 +3 & k=1 \\
9 h^3 -3 & k=2 \\
\end{array}
\right. ,
\end{array} $$
which add up to $H^0(\EEE,3h\Phi)= (3h)^3$.\\
Now $\chi (hF - kM)=9 h^3+\frac{3}{2}k(3-k^2)$ by the Riemann- Roch theorem (cf.\ \eqref{formula: Riemann Roch Z}): we notice that $dim (N_{h,k})$ equals $\chi (hF - kM)$ for $k=0,1,2$, $h>0$. This generalizes the result of Remark \ref{rem: F-aM and eigenspaces} and allows one to describe projective models of $\mathcal{Z}$ obtained from the maps associated to the divisors $hF-kM$, for each $h>0$, $k=0,1,2$.  }\end{rem}
\subsection{The first eigenspace}\label{section: the first eigenspace}
We now analyze the projection to the eigenspace relative to the eigenvalue 1, i.e\ the map $m_0$ on $\EEE$ given by $ (x_1x_2x_3:x_1x_2y_3:x_1y_2x_3:x_1y_2y_3:y_1x_2x_3:y_1x_2y_3:y_1y_2x_3:y_1y_2y_3:z_1z_2z_3).$\\
Considering the coordinate functions of $m_0$, we observe that they are invariant not only under the action of $\phip$, but also under the action of $\phi:\left((x_1:y_1:z_1),(x_2:y_2: z_2),(x_3:y_3: z_3)\right)$ $\ra$ $\left((x_1:y_1:\zeta z_1)\right.,$ $(x_2:y_2:\zeta^2 z_2),$ $\left.(x_3:y_3: z_3)\right)$. It is easy to see that the map is $9:1$ on $\EEE$, and hence the image gives a model of the Calabi--Yau variety $\mathcal{Y}$ which desingularizes $\EEE/\langle \phi,\phip\rangle$. So $\mathcal{Z}$ is a $3:1$ cover of $m_0(\EEE)$. The Calabi--Yau $\mathcal{Y}$, of which $m_0(\EEE)$ is a birational model, is still interesting, so we describe the map $m_0$ in some details. In this section we prove the following:
\begin{prop}\label{prop: m_0} The map $m_0:\EEE\ra\mathbb{P}^{8}$ is well defined on $\EEE$, and is a $9:1$ map on its image. Its differential fails to be injective only on the curves $C^{i}_{j,k}$ (cf.\ Example \ref{ex: curves Cijk and Aijk}).\\
The variety $m_0(\EEE)$ is a $3:1$ cover of $\sigma(\mathbb{P}^1\times \mathbb{P}^1\times \mathbb{P}^1)$ where $\sigma:\mathbb{P}^1\times \mathbb{P}^1\times \mathbb{P}^1\ra \mathbb{P}^7$ is the Segre embedding. Moreover $m_0(\EEE)$ is contained in the Fermat cubic hypersurface in $\mathbb{P}^8$.\\
The map $m_0$ induces the $3:1$ map $m_F:\mathcal{Z}\ra\mathbb{P}^8$ associated to the nef and big divisor $F$.\end{prop}
It is immediate to check that the map $m_0$ is $9:1$. To analyze its differential, we first consider $m_0$ as defined on $\mathbb{P}^2\times \mathbb{P}^2\times \mathbb{P}^2$ and then we will restrict it to $\EEE$. We recall that $\mathbb{P}^2$ is covered by its open subsets $U_x:=\{(x:y:z)|x\neq 0\}$, $U_y$ and $U_z$. Since the point $(0:0:1)\not\in E$, it suffices to consider the open sets $U_x$ and $U_y$, but the map is totally symmetric in the $x_i$ and $y_i$, so it is enough to study the map on the open set $U_x  \times U_x \times U_x $ of $\mathbb{P}^2\times \mathbb{P}^2\times \mathbb{P}^2$:
{\renewcommand{\arraystretch}{1.6}
$$
\begin{array}{cccccll}
U_x & \times & U_x & \times & U_x & \lra & U \subset \C^8 \\
(y_1,z_1) & \times & (y_2,z_2) & \times & (y_3,z_3) & \lmt &  \big( y_3 , y_2 , y_2 y_3 , y_1, y_1 y_3 , y_1 y_2 , y_1 y_2 y_3 , z_1 z_2 z_3 \big) \, .
\end{array} 
$$ }
The Jacobian is given by
\begin{equation*} 
J_{m_0} =\left(
\begin{array}{llllll}
 0 & 0 & 0 & 0 & 1 & 0 \\
 0 & 0 & 1 & 0 & 0 & 0 \\
 0 & 0 & y_3 & 0 & y_2 & 0 \\
 1 & 0 & 0 & 0 & 0 & 0 \\
 y_3  & 0 &  0  & 0 & y_1 & 0 \\
 y_2 &  0  &  y_1  & 0 & 0 & 0 \\
 y_2 y_3 & 0 & y_1 y_3 & 0 & y_1 y_2 & 0 \\
 0 &  z_2 z_3 & 0 & z_1 z_3 & 0 & z_1   z_2 
\end{array}
\right)
\end{equation*} 
Now we restrict our attention to the tangent space to $\EEE$: the tangent vectors $(u,v)$ to $E$ in $(y,z)$ satisfy $u \frac{\partial f}{\partial y} + v  \frac{\partial f}{\partial z} = 0$, where $f= 1 + y^3 + z^3$ is the equation of $E$ in $U_x$, thus $(u,v)=\lambda (- \frac{\partial f}{\partial z} , \frac{\partial f}{\partial y} ) =\lambda (- 3 z^2, 3 y^2)$, $\lambda\in \C$.
Hence tangent vectors of $\EEE$ in the point $q:=(y_1,z_1, y_2,z_2,y_3, z_3 )$ are\\ $(u_1,v_1, u_2,v_2,u_3, v_3) = (- \frac{\partial f_1}{\partial z_1} \lambda, \frac{\partial f_1}{\partial y_1} \lambda,  - \frac{\partial f_2}{\partial z_2} \mu , \frac{\partial f_2}{\partial y_2} \mu , - \frac{\partial f_3}{\partial z_3} \rho, \frac{\partial f_3}{\partial y_3} \rho)$, where $(\lambda,\mu,\rho) \in \C^3 \simeq T_q \EEE$. \\
The Jacobian fails to be injective where $J_{m_0}(u_1,v_1, u_2,v_2,u_3, v_3)^t=0$, which gives the following equations: 
$$\left\{\begin{array}{ccc}- \frac{\partial f_3}{\partial z_3} \rho& =& 0\\
- \frac{\partial f_2}{\partial z_2} \mu &=& 0\\
- \frac{\partial f_1}{\partial z_1} \lambda &=& 0\\ 
z_2 z_3 \frac{\partial f_1}{\partial y_1} \lambda + z_1 z_3 \frac{\partial f_2}{\partial y_2} \mu +  z_1   z_2  \frac{\partial f_3}{\partial y_3} \rho &=& 0
\end{array}\right.\Ra\left\{\begin{array}{ccc}z_3 =0\ (\Ra y_3 = \zeta^c \neq 0)& or &\rho = 0\\
z_2 =0\ (\Ra y_2 = \zeta^b \neq 0)& or &\mu = 0\\
z_1 =0\ (\Ra y_1 = \zeta^a \neq 0)& or &\lambda = 0\end{array}\right.
$$
Thus if either $z_i \neq 0 \, \forall i $ or $z_i =0, \, z_j,z_k \neq 0 \mbox{ for } \{i,j,k\}=\{1,2,3\} $, then $ \lambda = \mu = \rho = 0$ which gives no points where $(J_{m_0})_{|\EEE}$ is not injective. On the other hand the condition $z_1 \neq 0, \, z_j = 0 \mbox{ for } j \neq 1 $ (resp. $z_i = 0 \,\ \forall i $) implies $\lambda = 0$, but  does not give conditions on $\mu, \rho$ (resp. $\lambda, \mu, \rho$). We recall that the condition $z_i=0$ gives exactly the fixed points $p_i$ on the $i$-th copy of $E$ in $\EEE$. 
Therefore the map $(J_{m_0})_{|\EEE}$, and also $m_0$, fails to be injective on the curves $C^k_{i,j}$ (and in particular at the fixed points $p_i\times p_j\times p_k$). The reason is that these curves are invariant not only under the action of $\phip$, but also under $\phi$.

The map $m_0$ can also be described in a different way, which exhibits $m_0(\EEE)$ as $3:1$ cover of a subvariety in $\mathbb{P}^7$.  Consider the composition $\gamma$ of the projection $\alpha:\EEE\ra\mathbb{P}^1$ of each elliptic curve $E\subset\mathbb{P}^2$ on the first two coordinates and the Segre embedding $\sigma:(\mathbb{P}^1)^3\ra \mathbb{P}^7$
$$\begin{array}{ccccccccc} \gamma:&\EEE&\stackrel{\alpha}{\ra} &\mathbb{P}^1\ \ \times& \mathbb{P}^1\ \ \times&\mathbb{P}^1&\stackrel{\sigma}{\ra}&\mathbb{P}^7\\&
x_i^3+y_i^3+z_i^3=0&\mapsto &((x_1:y_1),&(x_2:y_2),&(x_1:y_1))&\mapsto&(x_1x_2x_3:x_1x_2y_3:\ldots:y_1y_2y_3)\end{array}$$ The Segre map $\sigma$ is well known to be an embedding and the map $\alpha$ is clearly $3^3:1$, hence $\gamma$ is $3^3:1$.\\ 
Let us denote by $X_0,\ldots X_8$ the coordinates on the target projective space of the map $m_0$. The map $\gamma$ is the composition of projection of $m_0$ with the projection on the hyperplane $\mathbb{P}^7\subset\mathbb{P}^8$ with coordinates $X_0,\ldots, X_7$. 
Thanks to this description one can show that $m_0(\EEE)$ is contained in certain quadrics and a cubic hypersurface. Indeed
the variety $\sigma(\PP^1\times\PP^1\times\PP^1)$ is contained in the quadrics
$$ \begin{array}{ccccc} 
X_0 X_3 = X_1 X_2, & X_0 X_5 = X_1 X_4, & X_0 X_7 = X_2 X_5, & 
X_0 X_7 = X_1 X_6, & X_0 X_6 = X_2 X_4, \\ X_0 X_7 = X_3 X_4, &
X_1 X_7 = X_3 X_5, & X_2 X_7 = X_3 X_6, & X_4 X_7 = X_5 X_6
\end{array} $$ 
and since $x_i^3 + y_i^3 + z_i^3 = 0$,we have $-(X_8)^3 = -(z_1 z_2 z_3)^3 = -(-x_1^3 -y_1^3)(-x_2^3 -y_2^3)(-x_3^3 -y_3^3) = \\X_0^3 + X_1^3 + X_2^3 + X_3^3 + X_4^3 + X_5^3+ X_6^3+ X_7^3$, so $m_0(\EEE)$ is contained in the Fermat cubic in $\mathbb{P}^8$, $F_8:=V(\sum_{i=0}^8X_i^3)$.\\
It is now clear that the projection $(X_0:\ldots:X_8)\ra (X_0:\ldots X_7)$ restricted to $F_8$ and to $m_0(\EEE)$ is a cyclic $3:1$ map with cover transformation $(X_0:\ldots: X_7:X_8)\ra (X_0:\ldots :X_7:\zeta X_8)$.

\begin{rem}\label{rem: m_0 big and nef}{\rm The map $m_F$ induced by $m_0$ on $\mathcal{Z}$ does not contract curves. This guarantees that $F$ is a big and nef divisor, indeed $F^3>0$ (i.e\ $F$ is big) and for each curve $C\in \mathcal{Z}$,  $FC=3deg(m_F(C))>0$, and by \cite[Theorem 1.26]{De} this suffices to conclude that $F$ is nef.}\end{rem}

The inclusion $m_0(\EEE)\subset F_8$ is interesting in view of the paper \cite{CDP} where the authors suggest that a generalized mirror for the Calabi--Yau 3-fold $\mathcal{Z}$ is a quotient of the Fermat cubic in $\mathbb{P}^8$ by an automorphism of order 3. Here we proved that there exists a $3:1$ map from $\mathcal{Z}$ to a singular model of the Calabi--Yau variety $\mathcal{Y}=\widetilde{\mathcal{Z}/\ZZZ}$ which is contained in this cubic in $\mathbb{P}^8$. The Hodge numbers of $\mathcal{Y}$ are $h^{1,1}=84$ and $h^{2,1}=0$.\\
In \cite{CDP} the authors observe that the middle cohomology of the desingularization $\widetilde{F_8/G}$ of the quotient of $F_8$ by certain groups $G$ has the following Hodge numbers: \begin{equation}\label{eq: Hodge diamond Candelas}H^7:\ \ \ 0\ \ 0\ \ 1\ \ \beta\ \ \beta\ \ 1\ \ 0\ \ 0.\end{equation} 
For a certain choice of the action of the group $G\simeq \ZZZ$, the value of $\beta$ is $36$ and hence $h^{4,3}(F_8/G)=h^{1,1}(\mathcal{Z})$. The space $H^{4,3}(\widetilde{F_8/G})$ is the complex moduli space of $\widetilde{F_8/G}$ and has the same dimension of the K\"ahler moduli space of $\mathcal{Z}$. Requiring that the dimension of the complex moduli space of a variety coincides with the dimension of the K\"ahler moduli space of another variety is one of the necessary conditions for the two varieties to be mirrors. In \cite{CDP} deeper relations between the complex moduli  of $\widetilde{F_8/G}$ and the K\"ahler moduli of $\mathcal{Z}$ are found using the Yukawa coupling. Because of this the authors suggest that $\widetilde{F_8/G}$ could be a ``generalized mirror" of $\mathcal{Z}$.\\
Now we observe that, if $G$ is trivial, then $\beta$ in \eqref{eq: Hodge diamond Candelas} is 84 and $h^{4,3}(F_8)=h^{1,1}(\mathcal{Y})$. Thus we observe that $F_8$ has the Hodge numbers of the generalized mirror of $\mathcal{Y}$ (a desingularization of $\mathcal{Z}/\ZZZ$). This was already noticed in \cite[Section 6.1.3]{KLS}, where the authors analyze a deeper relation between $\mathcal{Y}$ and $F_8$ based on their $L$-functions (cf.\ \cite[Theorem 2]{KLS}).\\ 
We observe that in these two generalized mirrors the desingularization of a quotient by $\ZZZ$ is involved:
$$\begin{array}{|c|c|}  
\hline\mbox{CY} &\mbox{conjectured generalized mirror}\\
\hline 
\mathcal{Z}&\widetilde{F_8/\ZZZ}\\
\hline
\mathcal{Y}=\widetilde{\mathcal{Z}/\ZZZ}&F_8\\\hline\end{array}$$
The fact that the Calabi--Yau variety $\mathcal{Y}$ admits a birational model inside the variety $\sum_{i=0}^8X_i^3=0$ (and $\mathcal{Z}$ a $3:1$ map to a subvariety of $\sum_{i=0}^8X_i^3=0$) could be useful to give a geometric explanation of the relations between $\mathcal{Z}$ and its generalized mirror and between $\mathcal{Y}$ and $F_8$.

\subsection{The second eigenspace}\label{section: second eigenspace} 
We now analyze the projection on the eigenspace of the eigenvalue $\zeta$, i.e\ the map $m_1$ on $\EEE$ given by $(z_1 x_2 x_3: z_1 x_2 y_3 : z_1 y_2 x_3 : z_1 y_2 y_3 : x_1 z_2 x_3 : x_1 z_2  y_3 : y_1 z_2 x_3 : y_1 z_2 y_3 : x_1 x_2 z_3 : x_1 y_2 z_3 : y_1 x_2 z_3 : y_1 y_2 z_3 )$. We summarize the properties of this map in the following Proposition, which is proved in this section:
\begin{prop}\label{prop: m_1} The base locus of the map $m_1:\EEE\ra\mathbb{P}^{11}$ consists of the 27 points $p_{i,j,k}$. The map $m_1$ contracts the 27 curves $C^i_{j,k}$ and is $3:1$ on $\EEE$  away from these curves. Its differential is injective away from the 27 contracted curves. The image $m_1(\EEE)$ has 27 singular points, the images of the curves $C^i_{j,k}$ , which are ordinary double points.\\
The map $m_1$ induces a well defined map over $\widetilde{\EEE}$ which sends the 27 exceptional divisors $\exijk$ to 27 copies of $\mathbb{P}^2$ and whose differential is injective away from the contracted curves.\\ 
The map $m_1$ induces the map $f_{F-M}$ on $\mathcal{Z}$ associated to the nef and big divisor $F-M$. It contracts the curves $A^i_{j,k}$ and is the isomorphism $\mathcal{Z}-\bigcup_{i,j,k=1}^3A^i_{j,k}\ra f_{F-M}\left(\mathcal{Z}-\bigcup_{i,j,k=1}^3A^i_{j,k}\right)$ away from the contracted curves.\end{prop}
By the definition of $m_1$, it is clear that the base locus is given by the condition $z_1=z_2=z_3=0$ and hence the base locus consists of the 27 points $p_{i,j,k}$. Let $q$ be one of the following 27 points $(0: \dots : 0 : 1 : - \zeta^b  :  - \zeta^a  :  \zeta^{a + b})$,  $(0: \dots : 0: 1 : - \zeta^c : -\zeta^a :  \zeta^{a+c}:0: \dots : 0)$ and $ (1: -\zeta^c :  -\zeta^b : \zeta^{b+c}: 0 : \dots : 0)$. Then the inverse image of $q$ is a curve $C^i_{j,k}$ (for example $C^1_{1,1}$ is sent to $(1:-\zeta:-\zeta:\zeta^2:0:\ldots:0)$). The inverse image of all the other points in $m_1(\EEE)$ consists of 3 points, so the map is generically $3:1$.\\
To study the Jacobian we consider the open subset $U_z  \times U_x \times U_x $ (where the map is surely defined, because the base locus is defined by $z_1=z_2=z_3=0$).
Since the tangent vectors to $\EEE$ in $(x_1, y_1, y_2, z_2, y_3, z_3 )$ are $(u_1,v_1, u_2,v_2,u_3, v_3) = (- \frac{\partial f_1}{\partial y_1} \lambda, \frac{\partial f_1}{\partial x_1} \lambda,  - \frac{\partial f_2}{\partial z_2} \mu , \frac{\partial f_2}{\partial y_2} \mu , - \frac{\partial f_3}{\partial z_3} \rho, \frac{\partial f_3}{\partial y_3} \rho)$, where $(\lambda,\mu,\rho) \in \C^3 \simeq T_p \EEE$, the restriction of the differential of $m_1$ to the tangent space of $\EEE$ has kernel:
$$\left\{\begin{array}{ccc}- \frac{\partial f_3}{\partial z_3} \rho &= &0\\ - \frac{\partial f_2}{\partial z_2} \mu &=& 0\\ - z_2 \frac{\partial f_1}{\partial y_1} \lambda + x_1 \frac{\partial f_2}{\partial y_2} \mu& =& 0 \\ - z_2 \frac{\partial f_1}{\partial x_1} \lambda + y_1 \frac{\partial f_2}{\partial y_2} \mu &=& 0\\
- z_3 \frac{\partial f_1}{\partial y_1} \lambda + x_1  \frac{\partial f_3}{\partial y_3} \rho& =& 0 \\
- z_3 \frac{\partial f_1}{\partial x_1} \lambda + y_1 \frac{\partial f_3}{\partial y_3} \rho &=& 0 
\end{array}
\right.\Ra\left\{\begin{array}{ccc}z_3 =0\ (\Ra y_3 = \zeta^c \neq 0)& or &\rho = 0\\ z_2 =0\ (\Ra y_2 = \zeta^b \neq 0)& or &\mu = 0\\- z_2 y_1^2 \lambda + x_1 y_2^2 \mu = 0\\- z_2 x_1^2 \lambda + y_1 y_2^2 \mu = 0\\- z_3 y_1^2 \lambda + x_1 y_3^2 \rho = 0 \\- z_3 x_1^2 \lambda + y_1 y_3^2 \rho = 0
\end{array}\right.
$$
Thus, for $z_2, z_3 \neq 0 \Ra \mu = \rho = 0$ and, considering for example the third and the fourth equation, we obtain
$$\left\{\begin{array}{cc}
- z_2 y_1^2  \lambda = 0\\
- z_2 x_1^2  \lambda = 0
\end{array} \right. \Ra \lambda = 0 \mbox{ (since $x_1$, $y_1$ cannot be both zero),}$$
which implies $\lambda=\rho=\mu=0$. This condition gives no points where the differential is not injective. Similarly, if
 $z_2 = 0, \, z_3 \neq 0 $, then $\rho = 0$ and (by the previous argument) $\lambda = \mu = 0$.
On the contrary in case  $z_2 = z_3 = 0$ one obtains  $\mu = \rho = 0$, but no conditions on $\lambda$, which corresponds to curves where the differential is not injective. The curves  $C^i_{j,k}$ are contracted to 27 singular points of $m_1(\EEE)$, which are the only singular points of the image.\\ 
In order to prove that these singular points are ordinary double points we consider some relations among the coordinate functions $N_i$, for $i=0, \ldots, 11$, of $m_1$. There are $15$ quadratic relations involving these monomials which are induced by the Segre embedding:
$$\begin{array}{l|l} 
\mathcal{Q}_i:=V(N_0N_{2i+1}-N_1N_{2i}),\  i=1,2,3&
\mathcal{Q}_i:=V(N_0N_{2i+1}-N_2N_{2i}),\  i=4,5\\
\mathcal{Q}_i:=V(N_1N_{2i-3}-N_3N_{2(i-2)}),\  i=6,7&
\mathcal{Q}_i:=V(N_2N_{2i-11}-N_3N_{2(i-6)}),\ i=8,9\\
\mathcal{Q}_{10}:=V(N_4 N_7 - N_5 N_6)&
\mathcal{Q}_{i}:=V(N_4N_{i-1}-N_6N_{i-3}),\  i=11,12\\
\mathcal{Q}_{i}:=V(N_5N_{i-3}-N_7N_{i-5}),\ i=13,14&
\mathcal{Q}_{15}:=V(N_8 N_{11} - N_9 N_{10})
\end{array}$$

We observe that, due to $x_i^3 + y_i^3 + z_i^3 = 0$, $i= 1,2,3$,
we obtain also $6$ cubic equations relating the monomials:
\small
$$\begin{array}{c|c|c}
\mathcal{C}_1:=V(N_0^3 + N_1^3 - (N_8^3 + N_{10}^3)) &
\mathcal{C}_2:=V(N_2^3 + N_3^3 - (N_9^3 + N_{11}^3))&
\mathcal{C}_3:=V(N_0^3 + N_2^3 - (N_4^3 + N_6^3)) \\
\mathcal{C}_4:=V(N_1^3 + N_3^3 - (N_5^3 + N_7^3))&
\mathcal{C}_5:=V(N_4^3 + N_5^3 - (N_8^3 + N_9^3))&
\mathcal{C}_6:=V(N_6^3 + N_7^3 - (N_{10}^3 + N_{11}^3)) 
\end{array} $$
\normalsize
So $m_1(\EEE)$ is contained in the intersection of all the varieties defined by these equations. As done before one can compute the Jacobian of these relations and analyze it at the singular points (for example at the point $p:= (0: \ldots : 0 : 1 : - 1 :  - 1 :  1)$ which is image of the curve $C^3_{3,3}$). We get $7$ independent relations among the $12$ variables $N_i$, for example choosing 
$\mathcal{Q}_4$, $\mathcal{Q}_6$, $\mathcal{Q}_{10}$, $\mathcal{Q}_{13}$, $\mathcal{C}_1$, $\mathcal{C}_2$ and $\mathcal{C}_3$. Hence $dim (ker J_p) = 12 - 7 = 5$ and $ker J_p = <(a: b : -a : -b: c :d : -c : -d : e : -e :-e : e)>$, $a,b,c,d,e\in \C$.
Once we projectivize this yields
$T_p (\mathcal{Q}_4,\ \mathcal{Q}_6,\ \mathcal{Q}_{10},\ \mathcal{Q}_{13},\ \mathcal{C}_1,\ \mathcal{C}_2,\ \mathcal{C}_3) \simeq \PP^4_{(a:b:c:d:e)}.$
In the affine coordinate chart $N_{11}= 1$ the tangent vectors are given by $(a, b, c, d)$.  Considering the quadric $\mathcal{Q}_2$ we obtain that $ab-cd = 0.$ Thus the points we obtain by contracting one of the $27$ curves are ordinary double points.\\
We already said that the map $m_1$ is not defined in the 27 points $p_{i,j,k}$ $i,j,k=1,2,3$, and hence we consider the blow up, $\widetilde{\EEE}$, of $\EEE$ in the base locus of $m_1$. As in Section \ref{section: the first eigenspace} we consider the map $m_1$ extended to $(\mathbb{P}^2)^3$ and we restrict it to the open subset $U_x \times U_x \times U_x$ which obviously contains the 27 points we wish to examine:
$$
\begin{array}{cccclll}
U_x & \times & U_x &\times & U_x & \stackrel{m_1}{\dashrightarrow} & \PP^{11} \\
(y_1,z_1) & \times & (y_2,z_2) & \times & (y_3,z_3) & \lmt & (z_1: z_1 y_3 : z_1 y_2 : z_1 y_2 y_3 : z_2 : z_2 y_3 : z_2 y_1 : z_2 y_1 y_3 : \\ 
&&&&&& z_3 : z_3 y_2 : z_3 y_1 : z_3 y_1 y_2) \, .
\end{array} $$
In $U_x \times U_x \times U_x$ the fixed points are given by $p_{i,j,k}=\left( (-\zeta^i,0), (-\zeta^j,0), (-\zeta^k,0) \right)$. We study what happens locally when we approach the point $p_{i,j,k}$. We consider the parametrized line through $p=p(0)$, $p(t) = \left( (-\zeta^i+t u_1,tv_1), (-\zeta^j+t u_2,tv_2), (-\zeta^k+t u_3,tv_3) \right)$, 
where $\left( (u_1,v_1),(u_2,v_2),(u_3,v_3) \right) \in (\C^2)^3$ and $t \in \C$. Under $m_1$ the point $p(t)$ is mapped to
$$ \begin{array}{rl}
m_1(p(t))= & (t v_1: t v_1 (-\zeta^k+t u_3) : t v_1 (-\zeta^2+t u_2) : t v_1 (-\zeta^j+t u_2) (-\zeta^k+t u_3) : \\ 
& t v_2 : t v_2 (-\zeta^k+t u_3) : t v_2 (-\zeta^i+t u_1) : t v_2 (-\zeta^i+t u_1) (-\zeta^k+t u_3) : \\
& t v_3 : t v_3 (-\zeta^j+t u_2) : t v_3 (-\zeta^i+t u_1) : t v_3 (-\zeta^i+t u_1) (-\zeta^j+t u_2)).
\end{array} $$
Let $E_x:=E\cap U_x$.  The  coordinates on $\mathcal{T}_p(E_x \times E_x \times E_x)$ are $v_1,v_2,v_3$, since $u_1=u_2=u_3=0$, and thus the exceptional divisor of $\widetilde{\EEE}$ over $p$ is mapped to a $\PP^2 =\PP(\mathcal{T}_p\EEE) =\PP^2_{(v_1:v_2:v_3)}$ linearly embedded in $\mathbb{P}^{11}$. It remains to prove that the differential of the map $m_1:\widetilde{\EEE}\ra \mathbb{P}^{11}$ is injective on the exceptional divisors. To do this let us choose as complex coordinates of $\EEE$ on $E_x\times E_x\times E_x$ the coordinates $z_i$. So $\EEE$ is locally isomorphic to $\C^3_{(z_1,z_2,z_3)}$ and $z_i=\sqrt[3]{-1-y_i^3}$. Blowing up $\C^3_{(z_1,z_2,z_3)}$ in $(0,0,0)$ we obtain a variety locally isomorphic to $\mathbb{C}_{(z_1,b,c)}$, where $z_2=bz_1$ and $z_3=cz_1$. The action of $\widetilde{\phip}$ on the coordinates $(z_1,b,c)$ is $(z_1,b,c)\mapsto (\zeta z_1,b,c)$. Hence the quotient $\mathbb{C}^3_{(z_1,b,c)}/\phip$ is locally isomorphic to a copy of $\mathbb{C}^3$ with coordinates $(z_1^3,b,c)$. Computing the Jacobian of the map induced by $m_1$ on this quotient, one finds that the rank of the Jacobian is 3 (i.e\ is maximal), hence the map induced by $m_1$ on $\widetilde{\EEE}$ has an injective differential (except on the contracted curves).

\begin{rem}{\rm 
As in the case of $m_0$, one proves that $F-M$ is big and nef. Indeed $m_1$ contracts the curves $A^k_{i,j}$ and $(F-M)A^k_{i,j}=0$. All the non contracted curves have positive intersection with $F-M$ (cf.\ Remark \ref{rem: m_0 big and nef}). Hence for each curve $C$ in $\mathcal{Z}$, $(F-M)C\geq 0$, so $F-M$ is nef. Since $(F-M)^3>0$, it is also big.  In particular $h^i(F-M)=0$, $i>0$.}\end{rem}
We observe that using the intersection form on $Pic(\mathcal{Z})$ one immediately finds that the image of the divisor $B_{i,j,k}$ under $m_1$ is a linear subspace (we computed explicitly this result on the blow up), indeed $(F-M)^2B_{i,jk}=1$. 

\subsection{The third eigenspace}\label{sec: third eigenspace}
We now analyze the projection on the eigenspace of the eigenvalue $\zeta^2$, i.e\ the map $m_2$ on $\EEE$ given by $ (x_1z_2z_3:y_1z_2z_3:z_1z_2x_3:z_1z_2y_3:z_1x_2z_3:z_1y_2z_3).$ We summarize the properties of this map in the following proposition, which is proved in this section. Let us denote by $S_{j}^1$ (resp. $S_{j}^2$, $S_{j}^3$) the surface $p_j\times E\times E$ (resp. $E\times p_j\times E$, $E\times E\times p_j$).
\begin{prop}\label{prop: m_2} The base locus of the map $m_2:\EEE\ra\mathbb{P}^{11}$ consists of the 27 curves $C^i_{j,k}$. The map $m_2$ contracts the 9 surfaces $S^{i}_j$ and is $3:1$ on $\EEE$ away from these surfaces. Its differential is injective away from the 9 contracted surfaces. 
The map $m_2$ is a $3:1$ dominant rational map between $\EEE$ and the desingularization $\widetilde{V_{3,3}}$ of the threefold  \begin{equation}\label{formula: V33}V_{3,3}:=\left\{\begin{array}{rrrrrrr}-X_0^3&-X_1^3&+X_2^3&+X_3^3&&&=0\\&&X_2^3&+X_3^3&-X_4^3&-X_5^3&=0\end{array}\right.\mbox{(cf.\ \cite{Ki})}.\end{equation}
Let $\widetilde{\widetilde{\EEE}}$ be the blow up of $\widetilde{\EEE}$ along the curves $\widetilde{C^i_{j,k}}$. The map $m_2$ induces a map $\widetilde{\widetilde{\EEE}}\ra V_{3,3}$, defined everywhere, which sends the strict transform of the exceptional divisors $\exijk$ to the 27 linear subspaces $\mathbb{P}^2_{(s:t:u)}\simeq(-s:\zeta^is:-t:\zeta^jt:-u:\zeta^ku)\subset V_{3,3}$ and the strict transform of the curves $\widetilde{C^i_{j,k}}$ to the 27 rational curves $(0:0:-\lambda:\zeta^i\lambda:-\mu:\mu\zeta^j)$, $(-\lambda:\zeta^i\lambda:0:0:-\mu:\mu\zeta^j)$, $(-\lambda:\zeta^i\lambda:-\mu:\mu\zeta^j:0:0)$.\\ The variety $V_{3,3}$ has 9 singular points of type $(3,3,3,3)$ which are the contractions of the strict transforms of the surfaces $S^{i}_j$ and whose tangent cone is the cone over the Fermat cubic Del Pezzo surface.\\
The map $m_2$ induces a generically $1:1$ map on $\mathcal{Z}$ associated to the divisor $F-2M$. It is not defined on the curves $A^i_{j,k}$ and contracts the surfaces $\pi(\widetilde{S^{i}_j})$, where $\widetilde{S^{i}_j}$ is the strict transform of $S^{i}_j$ over $\widetilde{\EEE}$.\end{prop}
By the definition of $m_2$, it is clear that the base locus is given by the condition $z_i=z_j=0$, $i,j=1,2,3$, $i\neq j$ and hence the base locus consists of the 27 curves $C^i_{j,k}$.  Moreover one immediately sees that the surfaces with $z_i=0$, $i=1,2,3$, are contracted to points by the map $m_2$. Under the condition $z_i\neq 0$, for each $i\in\{1,2,3\}$, $m_2$ is $3:1$. As in Sections \ref{section: the first eigenspace}, \ref{section: second eigenspace} one can prove that the Jacobian $J_{m_2}$ restricted to $\EEE$ is injective away from the contracted surfaces $S^i_j$.\\ 
In \cite{Ki} a dominant rational map of degree three between $\EEE$ and the threefold $V_{3,3}$ is given. Comparing the map described in \cite{Ki} with $m_2$ one observes that they coincide, up to a choice of the signs of the coordinate functions of $\mathbb{P}^5$. In \cite{Ki}, Kimura observes that the existence of this map is predicted by the Tate conjecture. The variety $V_{3,3}$ has $9$ singularities of type $(3,3,3,3)$ at the points $(1 : -\zeta^a : 0:0:0:0 )$,  $(0: 0: 0:0: 1 : -\zeta^b )$ and $ ( 0:0: 1: -\zeta^c: 0:0)$ ($a,b,c=0,1,2$), which can be resolved by a simultaneous blow up. By a direct computation one shows that the tangent cone over the singularity is a Del Pezzo cubic in $\PP^3$ and more precisely the Fermat cubic in $\mathbb{P}^3$. We also observe that there are $27$ curves $(-s:\zeta^a s:-t:\zeta^b t:0:0), (0:0:-t:\zeta^b t:-u:\zeta^c u),  (-s:\zeta^a s:0:0:-u:\zeta^c u)$ and $27$ surfaces $(-s:\zeta^a s:-t:\zeta^b t:-u:\zeta^c u)$, $a,b,c=1,2,3$, on $V_{3,3}$ which contain the singular points and which do not appear in the image $m_2(\EEE)$. We will show that these curves and surfaces are contained in the image of $\widetilde{\widetilde{\EEE}}$.\\
It suffices to consider the problem locally: locally $\EEE$ is isomorphic to a copy of $\mathbb{C}^3_{(z_1,z_2,z_3)}$. Blowing up the origin one obtains the variety $V(bz_1-az_2)\cap V(cz_1-az_3)\cap V(cz_2-bz_3)\subset \mathbb{C}^3_{(z_1,z_2,z_3)}\times \mathbb{P}^2_{(a:b:c)}$, which is isomorphic to a copy of $\mathbb{C}^3$ with coordinates $(z_1,b,c)$, in the affine set $a=1$. Applying the map $m_2$ to these new coordinates $\left((x_1,y_1,z_1); (x_2,y_2,bz_1); (x_3,y_3,cz_1)\right)$ one obtains $(x_1bc:y_1bc:x_3b:y_3b:x_2c:y_2c)$. It is clear that, in the affine set $a=1$, the map is not defined on the curve $b=c=0$ and that the exceptional divisor $\mathbb{P}^2_{(a:b:c)}$ over the point $p_{i,j,k}$ identified by $x_h=1$, $h=1,2,3$, $y_1=-\zeta^i$, $y_2=-\zeta^j$, $y_3=\zeta^k$ is sent to $(-bc:\zeta^ibc:-ab:\zeta^jab:-ac:\zeta^jac)\simeq\mathbb{P}^2_{(ab:ac:bc)}$. The map restricted to the exceptional divisor $\mathbb{P}^2_{(a:b:c)}$ is $(a:b:c)\ra (bc:ac:ab)$, so it is a Cremona transformation.\\
Blowing up $\mathbb{C}^3_{(z_1,b,c)}$ along the curve $b=c=0$ one finds, as an open subset, a third copy of $\mathbb{C}^3$ with coordinates $(z_1, b, \gamma)$, related to the previous ones by $c=\gamma b$. Applying the map induced by $m_2$ one obtains the everywhere defined map on $\widetilde{\widetilde{\EEE}}$, locally given by $(x_1b: y_1b:x_2\gamma:y_2\gamma:x_3:y_3)$. Now one can directly check that the images of the strict transforms of $\exijk$ and of $\widetilde{C^i_{j,k}}$ are the ones given in the statement.  The intersection of the contracted surface $S^l_h$ with the strict transform of $\exijk$ is given by the lines contracted by the Cremona transformation induced by $m_2$ on $\exijk$.

\begin{rem}{\rm The divisor $F-2M$, associated to $m_2$, is a big divisor, but it is not nef, indeed $(F-2M)A^k_{i,j}<0$.}\end{rem}

The threefold $V_{3,3}$ is studied by several authors who give different models and descriptions of this threefolds. In \cite{WvG} it is proved that this Calabi--Yau is isomorphic to a $3^5:1$ cover of $\mathbb{P}^3$ branched along the configuration of six planes which was constructed in \cite{Hi}. The map associated to this cover is $\varsigma:(X_0:X_1:X_2:X_3:X_4:X_5)\ra (X_0^3-X_1^3:X_2^3-X_3^3:X_4^3-X_5^3:X_2^3+X_3^3)$, hence there exists a $3^6:1$ rational dominant map between $\EEE$ and $\mathbb{P}^3$, obtained by the composition $\varsigma\circ m_3$, defined by $\left((z_2z_3)^3(-x_1^3+y_1^3):(z_1z_2)^3(x_3^3-y_3^3):(z_1z_3)^3(-x_2^3+y_2^3):(z_1z_2)^3(x_3^3+y_3^3)\right)$.\\ Another construction of the same Calabi--Yau is given in \cite{Me} where it is shown that the $3^5:1$ map $V_{3,3}\dashrightarrow \mathbb{P}^3$ is the composition of a $3^4:1$ map $V_{3,3}\dashrightarrow T$ and a $3:1$ map $T\ra \mathbb{P}^3$, the last map being a $3:1$ cover of $\mathbb{P}^3$ branched along the six planes which are the faces of a cube.

\subsection{Other maps}\label{section: other maps}
In the previous section we analyzed the maps defined on $\mathcal{Z}$ associated to the divisors $F$, $F-M$ and $F-2M$. These divisors give $3:1$ or $1:1$ maps, but they do not give an image which is isomorphic to $\mathcal{Z}$. Indeed, the divisors $F$ and $F-M$ are big and nef, but not ample and $F-2M$ is not nef. In particular none of them is very ample. Here we want to prove that the divisor $2F-M$ is a very ample divisor and hence $f_{2F-M}(\mathcal{Z})$ is isomorphic to $\mathcal{Z}$, where $f_{2F-M}$ is the map defined by the divisor $2F-M$. In general it is not easy to find criteria which assure that a given divisor is very ample, and indeed we will prefer to consider explicitly the map associated to $2F-M$ and prove that it is an isomorphism on the image. However it is clear that since $F$ and $F-M$ are both nef and big, the sum of these divisors $2F - M$ is nef and big. \\
Let $m:\EEE\ra \mathbb{P}^N$ be the composition of the map $f_{2(3\Phi)}$ defined on $\EEE$ and the projection on the eigenspace of the eigenvalue $\zeta$ (with respect to the action of $\phip$). By Remark \ref{rem: hF-kM} we know that the map induced by $m$ on $\mathcal{Z}$ is the map $f_{2F-M}$. 
\begin{prop}\label{prop: 2F-M} The base locus of the map $m$ consists of the 27 points $p_{i,j,k}$. It induces an everywhere defined $3:1$ map over $\widetilde{\EEE}$, whose differential is injective. The map induced on $\mathcal{Z}$ is an isomorphism on the image and is associated to the divisor $2F-M$ which is very ample.\end{prop}
\proof By Remark \ref{rem: hF-kM} we have
$$ \begin{array}{rcl}
m:\EEE & \lra & \PP^{74}=\mathbb{P}(N_{2,1}) ,
\end{array}
$$
and we are considering monomials contained in the eigenspace relative to $\zeta$. \\
We consider only the monomials with one $z$, namely of the form 
$$v_iv_jv_k \mbox{ with } v_i \in \{z_ix_i,z_ix_i\},\  v_j \in\{x_j^2,x_jy_j,y_j^2\} \mbox{ and } v_k\in\{x_k^2,x_ky_k,y_k^2\},$$
where $\{i,j,k\} =\{1,2,3\}$. We restrict ourselves to the open subset defined by $x_h = 1$ for $h=1,2,3$. The rank of the Jacobian of $m$ is $3$. Thus the differential is injective where the map is defined.\\We recall that $F-M$ defines an isomorphism on $\mathcal{Z}$ except on the curves $A^i_{j,k}$, which are contracted, so $2F-M$ is an isomorphism outside the curves $A^i_{j,k}$ and thus it suffices to prove that $2F-M=F+F-M$ has an injective differential on these curves.  The blow up of $\EEE$ in the points $p_{i,j,k}$ is locally the blow up of $\mathbb{C}^3_{(z_1,z_2,z_3)}$ in the origin and is isomorphic to $\mathbb{C}^3_{(z_1,b,c)}$ on an open subset (cf.\ subsection \ref{sec: third eigenspace}). The quotient by the map $\widetilde{\phip}$ is locally isomorphic to $\mathbb{C}^3_{(z_1^3,b,c)}$. Applying the map induced by $m$ to $\mathbb{C}^3_{(z_1^3,b,c)}$ and computing the differential one obtains that it has rank 3 everywhere, in particular where $b=c=0$ (which is the curve corresponding to $A^1_{j,k}$). 
So the map $m$ induces on $\mathcal{Z}$ a map which is an isomorphism everywhere, thus divisor $2F-M$ is very ample.\eprf
\begin{rem}{\rm 
The divisor $F-M+F_1$ is a nef and big divisor on $\mathcal{Z}$. It is easy to check that it contracts the curves $A^2_{j,k}$ and $A^3_{j,k}$ (indeed it acts essentially as $F-M$ on these curves), but is an isomorphism on $A^1_{j,k}$ (indeed it acts on these curves as $2F-M$) and away from the curves $A^i_{j,k}$ (again because $F-M$ has the same property). One can also check that $(F-M+F_1)^2A^1_{j,k}=1$ and $(F-M+F_1)^2A^i_{j,k}=0$ if $i=2,3$. Analogously we have that $F-M+F_i$, $i=1,2,3$, contracts 18 curves among $A^h_{j,k}$ and is an isomorphism away from the contracted curves.\\
Similarly the map associated to the divisor $F-M+F_i+F_j$, $i\neq j$, $i,j\in\{1,2,3\}$, contracts 9 curves among $A^{h}_{k,l}$ and is an isomorphism away from these curves.}\end{rem}

\section{Other Calabi--Yau 3-folds}\label{sec: other CY 3-folds}
Until now we considered the well known Calabi--Yau 3-fold $\mathcal{Z}$. The aim of this section is to recall that, starting from the Abelian 3-fold $\EEE$ or from the Calabi--Yau 3-fold $\mathcal{Z}$, one can construct other Calabi--Yau 3-folds with different Hodge numbers and properties. We will describe some already known constructions and results. In Section \ref{sec: smoothing} we present two Calabi--Yau 3-folds obtained from a singular model of $\mathcal{Z}$ by a smoothing. One of these was unknown until now.
\subsection{Quotient by automorphisms}\label{section: quotient by automorphism} 
To construct $\mathcal{Z}$ we considered the desingularization of the quotient of the Abelian variety $\EEE$ by the automorphism $\phip$. Similarly we can consider quotients of $\EEE$ by other automorphisms, or quotients of $\mathcal{Z}$ by automorphisms induced by the ones of $\EEE$.

\begin{ex}{\rm We already saw an example of this construction in \ref{section: the first eigenspace}. Indeed the group $G\simeq (\ZZZ)^2=\langle \phip, \phi\rangle=\langle \zeta\times \zeta\times \zeta, \zeta\times \zeta^2\times 1\rangle$ acts on $\EEE$ and $\EEE/G$ is a singular 3-fold which admits a desingularization $\mathcal{Y}$ which is a Calabi--Yau. This 3-fold is already known (\cite{CH}, \cite{GvG}, \cite{Rohde}) and its Hodge numbers are computed in \cite{Rohde}: $h^{2,1}=0$, $h^{1,1}=84$. The automorphisms $\phip$ and $\phi$ commute on $\EEE$, this implies that $\phi$ induces an automorphism $\alpha$ of order 3 on $\EEE/\phip$ and also on $\mathcal{Z}$. Thus the 3-fold $\mathcal{Y}$ can be obtained also as desingularization of the quotient $\mathcal{Z}$ by the automorphism $\alpha$. The automorphism $\alpha$ preserves the 3-holomorphic form on $\mathcal{Z}$, and this guarantees that the quotient $\mathcal{Z}/\alpha$ has a desingularization which is a Calabi--Yau threefold.}\end{ex}
As in the previous example, one can construct automorphisms on $\mathcal{Z}$ considering the automorphisms of $\EEE$, which commute with $\phip$. These automorphisms induce automorphisms on $\mathcal{Z}$.\\ On each Abelian variety $A$, and in particular on $\EEE$, the translations by points are defined, indeed for each point $r\in A$ the map $t_r:A\ra A$ such that $t_r(q)=q+r$ for each $q\in A$ is an automorphism of $A$.  Let us assume that the order of $t_r$ is finite. The quotient $\EEE/t_r$ does not have the Hodge numbers of a Calabi--Yau variety, indeed translations preserve all the holomorphic forms on $\EEE$, thus $h^{1,0}(\EEE/t_r)=h^{2,0}(\EEE/t_r)=3 $. However the translation $t_r$ commutes with the automorphism $\phip$ if and only if $\phip(r)=r$. In this case the translation $t_r$ defines an automorphism on $\EEE/\phip$ and a desingularization of $\EEE/\langle \phip, t_r\rangle$ has the Hodge numbers of a Calabi--Yau 3-fold.\\
The Abelian variety $\EEE$ admits a larger group of automorphisms: $GL_3(\Z[\zeta])\subset Aut(\EEE)$. In particular, the automorphism $\phip\in GL_3(\Z[\zeta])$ is the diagonal matrix $(\zeta,\zeta,\zeta)$. Since all the matrices $M\in GL_3(\Z[\zeta])$ commute with the diagonal matrix (and in particular with $\phip$), the automorphisms of $\EEE$ given by the matrices $M$ induce automorphisms on $\mathcal{Z}$. Thus $GL_3(\Z[\zeta])\subset Aut(\mathcal{Z})$. If $\sigma\in SL_3(\Z[\zeta])$ is of finite order, then $\mathcal{Z}/\sigma$ has a desingularization which is a Calabi--Yau threefold. It is in general an open (and non trivial) problem to find explicitly such a desingularization and to compute  its Hodge numbers $h^{1,1}$, $h^{2,1}$.\\
In \cite{Oguiso 2} some quotients of $\EEE$ by subgroups of $SL_3(\Z[\zeta])$ and their crepant resolutions are analyzed (\cite[Theorem 3.4]{Oguiso 2}). We notice that the definition of Calabi--Yau variety in \cite{Oguiso 2} is slightly different from our definition, indeed in \cite{Oguiso 2} it is not required that $h^{2,0}=0$ and some particular singularities are admitted. Anyway, the 3-folds $X_{3,1}$ and $X_{3,2}$ in \cite[Theorem 3.4]{Oguiso 2} have $h^{2,0}=0$. \\
In \cite{Donten} the finite subgroups of $SL_3(\Z)$ are classified and the action of some of these subgroups on the product of three elliptic curves is studied. \\
In \cite{AW} certain Calabi--Yau varieties obtained as desingularization of quotients of an Abelian variety by a group of automorphisms are presented and their cohomology is computed.
\subsection{Elementary modifications}
In \cite{Fr} two constructions are considered, both of them are related to the presence of rational curves $C$ on a Calabi--Yau 3-fold $X$ such that the normal bundle of $C$ in $X$ is $\mathcal{N}_{C/X}\simeq \mathcal{O}_{\mathbb{P}^1}(-1)\oplus \mathcal{O}_{\mathbb{P}^1}(-1)$. The curves which satisfy this property are said to be of type $(-1,-1)$. The first construction is the elementary modification with respect to one of these curves, the second one is the smoothing of the threefold obtained by contracting these curves. We will see later that the curves $A^h_{i,j}$ on $\mathcal{Z}$ are rational curves of type $(-1,-1)$ and hence can be used for these constructions.\\
The elementary modification on a rational curve $C\subset X$ of type $(-1,-1)$ consists of a blow up of $C$ and a blow down: blowing up the curve $C$ on $X$, one obtains a variety $\hat{X}$ with an exceptional divisor $D$ which is isomorphic to $\mathbb{P}^1\times \mathbb{P}^1$ (which is a $\mathbb{P}^1$-bundle over the curve $C\simeq \mathbb{P}^1$). Now one can contract the first or the second copy of $\mathbb{P}^1$ in the exceptional divisor $\mathbb{P}^1\times \mathbb{P}^1$. One of these contractions is the opposite of the blow up $\hat{X}\ra X$, gives exactly $X$
and sends $D$ to $C\subset X$, the other one gives a new 3-fold $X'$ (in this case $D$ is sent to a rational curve $C'\subset X'$), which is said to be obtained by $X$ by an elementary modification (or a flop) on the curve $C$:
$$\begin{array}{ccccc}
X&\leftarrow &\hat{X}&\ra&X'\\
C&\leftarrow &D&\ra&C'\end{array}.$$
If $X$ is projective, $X'$ is not necessarily a projective variety, (see for example \cite[Examples 7.6, 7.7]{Fr}). In fact $X'$ may not even be K\"ahler.\\
In \cite{Oguiso} some fibrations on the 3-fold $\mathcal{Z}$ are described. They are induced by the projection of $\EEE$ on $E\times E$, the product of two factors of $\EEE$, and the general fiber is an elliptic curve. Thanks to the results in section \ref{section: the cohomology of Z} we conclude that these elliptic fibrations are associated to the divisors $F_i+F_j$ on $\mathcal{Z}$. These elliptic fibrations have some singular fibers which are the fibers over the points of $E\times E$ (the base of the fibration) which are fixed by $\phip$. These fibers consist of a rational curve and 3 planes. From this geometric description it is immediate that the rational curve is one of the curves $A^h_{i,j}$ and the planes are three of the $B_{i,j,k}$'s. More precisely, if the fibration considered is obtained from the projection $\EEE\ra E\times E$ on the second and the third factors, then the rational curves over the points $(p_i,p_j)\in E\times E$ are $A^1_{i,j}$ and the planes in the fibers are $B_{a,i,j}$, $a=1,2,3$. Once one fixes the projection $\EEE\ra E\times E$ the rational curves in the exceptional fibers depend only on $i,j$ and these curves are denoted by $l_{i,j}$ in \cite{Oguiso}. In \cite[Proposition 2.2]{Oguiso} it is proved that the curves $A^k_{i,j}$ (denoted by $l_{i,j}$ by Oguiso) are of type $(-1,-1)$. Moreover in \cite[Proposition 2.4]{Oguiso} the 3-folds $X_T$, obtained by elementary modifications on certain subsets $T$ of curves $A^k_{i,j}$, are proved to be Calabi--Yau varieties with $h^{2,1}=0$. By construction, the 3-folds $\mathcal{Z}$ and $X_T$ are birational, but in \cite[Theorem 3.1]{Oguiso 3} it is proved that they are not homeomorphic, i.e\ they are not equivalent from a topological point of view, but they are birationally equivalent.

\section{The smoothings}\label{sec: smoothing}
In this Section we construct a new Calabi--Yau threefold obtained by smoothing a singular model of $\mathcal{Z}$.\\ 
In \cite[Section 8]{Fr} the smoothing of a singular variety obtained by the contraction of certain curves on a 3-fold with trivial canonical bundle is analyzed. 
Let $V$ be a 3-fold with trivial canonical bundle and let $C_i$ be rational curves on it. Let $\overline{V}$ be the 3-fold obtained contracting the $C_i$ and let us assume that the $C_i$ contract to singular ordinary double points, $p_i$, on $\overline{V}$. Let $r:V\ra \overline{V}$ be the contraction. Let $V_t$ be the 3-fold obtained by smoothing $\overline{V}$, i.e\ $V_t$ is the fiber over $t$ of a proper flat map $f:\mathcal{V}\ra \Delta$, where $\Delta$ is the unit disc of $\C$, such that $f^{-1}(0)=\overline{V}$ and $f^{-1}(s)=V_s$ is smooth for each $s\in\Delta$, $s\neq 0$. Friedman proves the following results:
\begin{lemma}{\rm \cite[Lemma 8.7]{Fr}}\label{lemma: smoothing existence} Let $\overline{V}$ be a compact complex 3-fold with only ordinary double point singularities and let $\pi:V\ra \overline{V}$ be a small resolution such that the canonical bundle of $V$ is trivial. Let $p_i$ be the singularities of  $\overline{V}$ and $C_i$ be the curve $\pi^{-1}(p_i)$. Then there exists a first order deformation of $\overline{V}$ which is non trivial for the $p_i$ (i.e\ smooths them to first order) if and only if the fundamental classes $[C_i]$ in $H^{2}(V,\Omega_V^2)$ satisfy a relation $\sum_i\lambda_i[C_i]=0$ such that for every $i$, $\lambda_i\neq 0$.\end{lemma}
This Lemma gives a condition to assure that there exists a smoothing of the 3-fold $\overline{V}$. The following two lemmas describe the properties of this smoothing, if it exists.
\begin{lemma}\label{lemma: smoothing Hodge numbers}{\rm (\cite[Lemma 8.1]{Fr})}With the previous notations, let $e:\oplus_i\Z C_i\ra H_2(V,\Z)$ be the map which associates to each curve its class, then $H_2(V_t,\Z)$ is isomorphic to the cokernel of $e$ and $b_3(V_t)=b_3(V)+2s$, where $s$ is the rank of the kernel of $e$.\end{lemma}
\begin{lemma}\label{lemma: smoothing CY}{\rm (\cite[Lemma 8.2]{Fr})}With the previous notations: 1) if $h^{i,0}(V)=0$ for a certain $i$, then $h^{i,0}(V_t)=0$ for any small $t$.\\
2) If the canonical bundle of $V$ is trivial and $h^{1,0}(V)=0$, then the canonical bundle of $V_t$ is trivial.\end{lemma}
The previous lemma implies that, if $V$ is a Calabi--Yau variety, then $V_t$ is a Calabi--Yau variety. Moreover if one can describe the map $e$, its kernel and its cokernel, one immediately deduces $b_2(V_t)$ and $b_3(V_t)$. For a Calabi--Yau variety we have $b_2=h^{1,1}$ and $b_3=2+h^{2,1}+h^{1,2}=2+2h^{2,1}$. Thus knowing $e$, one determines the Hodge diamond of the Calabi--Yau variety $V_t$.\\
Now the idea is to apply these results to $V=\mathcal{Z}$ and to the contraction $m_1$ of the 27 curves $A_{i,j}^k$, $i,j,k=1,2,3$. Indeed $\overline{\mathcal{Z}}:=m_1(\mathcal{Z})$ is a 3-fold with only ordinary double points (this was proved in Proposition \ref{prop: m_1}, but is also the consequence of the cited result by Oguiso, \cite[Proposition 2.2]{Oguiso}, who proved that the curves $A^k_{i,j}$ are  of type $(-1,-1)$)  and $m_1:\mathcal{Z}\ra \overline{\mathcal{Z}}$ is the small resolution required by Lemma \ref{lemma: smoothing existence}. Now it suffices to show that there exists a relation in $H_2(\mathcal{Z},\Z)\simeq H^4(\mathcal{Z},\Z)$\begin{equation}\label{eq: condition smoothing 27}\sum_{i,j,k=1,2,3}\lambda_{i,j,k}A_{i,j}^k=0,\ \ \lambda_{i,j,k}\neq 0,\mbox{ for each }i,j,k=1,2,3,\end{equation} to prove that there exists a smoothing of $\overline{\mathcal{Z}}$. If this smoothing exists, then its fibers are Calabi--Yau 3-folds, by Lemma \ref{lemma: smoothing CY}. We described $A^k_{i,j}$ as linear combination of a basis of $H^{2,2}(\mathcal{Z})$ in \eqref{eq_ the curves Ahij}. Hence it suffices to substitute each $A_{i,j}^k$ with its expression in \eqref{eq: condition smoothing 27} and to determine $\lambda_{i,j,k}$ such that all the coefficients of the basis of $H^{2,2}(\mathcal{Z})$ are equal to zero. The following choice for the $\lambda_{i,j,k}$, all non zero, gives the relation:
\small $$4A^1_{1,1}+\sum_{i=2}^3 A^1_{1,i}+A^1_{2,1}-2\sum _{i=2}^3A^1_{2,i}+A^1_{3,1}-2\sum_{i=1}^3A^1_{3,i}-2\sum_{i=1}^3 A^2_{1,i}+\sum_{i=1}^3 A^2_{2,i}+\sum_{i=1}^3 A^2_{3,i}-2\sum_{i=1}^3 A^3_{1,i}+\sum_{i=1}^3 A^3_{2,i}+\sum_{i=1}^3 A^3_{3,i}=0.$$\normalsize
Hence there exists a smoothing of $\overline{\mathcal{Z}}$ and its fibers are Calabi--Yau varieties. We compute the Hodge numbers of these Calabi--Yau 3-folds, using Lemma \ref{lemma: smoothing Hodge numbers}. Indeed the map $e$ in this lemma is exactly the one described by the relations \eqref{eq_ the curves Ahij}. It is a trivial computation to show that its kernel has dimension 6 and hence we have that $b_3(V_t)=2+12=14$ and $b_2(V_t)=36-21=15$.
Thus the Hodge diamond of the Calabi--Yau varieties which are smooth fibers of the smoothing of $\overline{\mathcal{Z}}$ has the following Hodge diamond:
\begin{eqnarray}\label{eq: small Hodge numbers}\begin{array}{rrrrrrr}
&&&1&\\
&&0&&0\\
&0&&15&&0\\
1&&6&&6&&1\end{array}\end{eqnarray}
and Euler characteristic equal to $18$.
\begin{rem}{\rm  In \cite[Figure 1]{Braun:2009qy} the known Calabi--Yau threefolds with small Hodge numbers are listed. The Calabi--Yau threefold obtained as smoothing of $\overline{\mathcal{Z}}$ with Hodge diamond \eqref{eq: small Hodge numbers} is unknown.}\end{rem}

In a similar way one can consider the map $f_{F-M+F_3}$ associated to the divisor $F-M+F_3$. As proved in Section \ref{section: other maps}, this map contracts the 18 curves $A^k_{i,j}$, $i,j=1,2,3$, $k=1,2$ and is injective with an injective differential  away from these curves. The map $f_{F-M+F_3}:\mathcal{Z}\ra f_{F-M+F_3}(\mathcal{Z})$ contracts the curves $A^k_{i,j}$, $i,j=1,2,3$, $k=1,2$ to ordinary double points (indeed all these curves are of type $(-1,-1)$). Among these curves there exists the following relation:
$$2\sum_{i=1}^3A^1_{1,i}-\sum_{i=1}^3\left(A^1_{2,i}+A^1_{3,1}\right)-2\sum_{i=1}^3A^2_{1,i}+\sum_{i=1}^3\left(A^2_{2,i}+A^2_{3,1}\right)=0$$
thus one can smooth $f_{F-M+F_3}(\mathcal{Z})$ to Calabi--Yau varieties. By Lemma \ref{lemma: smoothing Hodge numbers} we compute the Hodge numbers of such a smooth Calabi--Yau: $b_3=2+4=6$, $b_2=36-16=20$. Thus the Hodge diamond is 
\begin{eqnarray}\label{eq: small Hodge numbers 2}\begin{array}{rrrrrrr}
&&&1&\\
&&0&&0\\
&0&&20&&0\\
1&&2&&2&&1\end{array}\end{eqnarray}
and Euler characteristic equal to $36$. \\
The situation is essentially the same considering the maps induced by $F-M+F_1$ and $F-M+F_2$.
\begin{rem}{\rm In \cite[Pag. 25, line 1 and Section 3.4.2]{Candelas:2008wb} a Calabi--Yau threefold with the Hodge diamond mirror of the one in \eqref{eq: small Hodge numbers 2} is given. In particular this Calabi--Yau is constructed as smooth quotient of a complete intersection. We do not know if this Calabi--Yau manifold is the mirror of the one constructed as smoothing of $f_{F-M+F_3}(\mathcal{Z})$ or if they only have specular Hodge numbers.}\end{rem}

In Section \ref{section: other maps} we noticed that the map associated to $F-M+F_1+F_2$ contracts the nine curves $A_{i,j}^3$, $i,j=1,2,3$, but we cannot use the map $f_{F-M+F_1+F_2}$ to construct a smoothing of $f_{F-M+F_1+F_2}(\mathcal{Z})$. Indeed it is immediate to check that the curves $A^3_{i,j}$, $i,j=1,2,3$, are linearly independent and hence there exists no relation among them as required by Lemma \ref{lemma: smoothing existence}, i.e\ there exists no smoothing of $f_{F-M+F_1+F_2}(\mathcal{Z})$.

\section{Conclusions}
In this paper we analyzed the geometry of the rigid Calabi--Yau threefold $\mathcal{Z}$. This Calabi--Yau is well known, but the careful description of its geometry allows us to give  a different interpretation of known results (for example of the generalized mirror) and to construct other Calabi--Yau threefolds.\\

The mirror conjecture says that two families of Calabi--Yau threefolds are mirror if the complex moduli space of one of them is locally isomorphic to the K\"ahler moduli space of the other one. 
The definition of the isomorphism among these two moduli spaces implies that the instanton corrections defined on $H^{1,1}$ of one of them can be reconstructed by the superpotential of the complex moduli defined on $H^{2,1}$ of the other Calabi--Yau threefold and thus involve the Yukawa coupling on the Calabi--Yau threefolds. The mirror conjecture is stated, and in particular cases proved, under two hypotheses on the families of Calabi--Yau threefolds: the families have to be at least 1--dimensional and have to admit a point with maximal unipotent monodromy. Families of Calabi--Yau threefolds without a point with maximal unipotent monodromy are described for example in \cite{Rohde}, \cite{GvG},\cite{G}. An extension of the mirror conjecture to these families is until now unknown. On the contrary, there are some ideas to extend the mirror conjecture to rigid Calabi--Yau threefolds (see \cite{Schimmrigk:1994ya}, \cite{Candelas:1990rm},  \cite{AG}). In particular in \cite{Candelas:1990rm} a generalized mirror for the threefold $\mathcal{Z}$ is proposed. In the classical mirror symmetry there often is a geometric link among families of Calabi--Yau threefolds and their mirrors and this link is very useful in the mirror construction: the most famous example is the mirror of the quintic in $\mathbb{P}^4$, which is the desingularization of the quotient of a particular quintic by a group of automorphisms. In this paper we have analyzed a lot of properties of the well known rigid Calabi--Yau threefold $\mathcal{Z}$ and we gave a geometric relation between it and its generalized mirror.\\
The other problem we analyzed is the construction of famillies of Calabi--Yau threefolds with given Hodge numbers. In \cite{Braun:2009qy} there is a list of the pairs of small numbers which can be the Hodge numbers of Calabi--Yau threefolds. Not all these pairs correspond to known Calabi--Yau threefolds. Here we constructed a new family with given Hodge numbers, starting from the Calabi--Yau threefold $\mathcal{Z}$.\\ 

In this paper we have obtained essentially three types of results: in sections \ref{section: cohomology of EEE} and \ref{sec: more on the trilinear form on Pic(EEE)} we described the trilinear form on $Pic(\mathcal{Z})$ and hence we gave information on the Yukawa coupling of $\mathcal{Z}$; in Section \ref{sec: projective models Z} we gave very explicit equations of projective models of $\mathcal{Z}$, related one of them with the generalized mirror of $\mathcal{Z}$ and suggested how to construct other models; in Section \ref{sec: smoothing} we constructed other Calabi--Yau threefolds starting from $\mathcal{Z}$. \\

In Section \ref{sec: EEE widetildeEEE Z and thier cohomology} the trilinear form on $Pic(\mathcal{Z})$ is computed. As we already said this gives information on the Yukawa coupling of $\mathcal{Z}$, but it is also important in view of the construction of the projective models of $\mathcal{Z}$. The computation of this form is very explicit and we essentially proved that the Picard group of $\mathcal{Z}$ splits in two parts, one of them comes from $Pic(\EEE)$, the other one from the resolution of the singularities of the quotient $\EEE/\phip$. The trilinear form on the second part is very elementary, the one on the first part depends only on the properties of the Abelian variety $\EEE$. We proved that the trilinear form on $Pic(\EEE)$ can be expressed in terms of determinants of certain matrices in $Mat_{3,3}(\Q[\zeta])$. This is useful because, of course, makes the computation easier, but could also have a deeper meaning, indeed this creates a strong relation between a significative part of the Yukawa coupling of $\mathcal{Z}$ and a group of matrices. This could help in clarifying the relation between string theory and exceptional supergravities, cf.\ \cite{Fer}.\\

The threefold $\mathcal{Z}$ is well known in the literature as desingularization of the quotient of an Abelian variety by a group of automorphisms, but there are not many explicit descriptions of its projective models. Here we provided three very explicit descriptions and suggested a strategy to obtain many others: We wrote down the equations of maps from $\EEE$ to projective spaces and related them with divisors on $\EEE$. Thanks to the strong relation between $\mathcal{Z}$ and $\EEE$ these maps give (singular) models of $\mathcal{Z}$. In particular, we analyzed three maps: The first one exhibits $\mathcal{Z}$ as a $3:1$ cover of another Calabi--Yau threefold $\mathcal{Y}$, which is contained in the cubic hypersurface $F_8$ in $\mathbb{P}^8$. This is interesting in view of the study of the mirror conjecture of rigid Calabi--Yau threefolds, indeed in \cite{Candelas:1990rm} a quotient of the cubic Fermat hypersurface $F_8$ is proved to be a generalized mirror of $\mathcal{Z}$ and the hypersurface $F_8$ is conjectured to be the generalized mirror of $\mathcal{Y}$ (see \cite{KLS}). Here we have provided a geometric relation between these two rigid Calabi--Yau threefolds ($\mathcal{Z}$, $\mathcal{Y}$) and their generalized mirrors. The second map gave a singular birational model of $\mathcal{Z}$. It is embedded in $\mathbb{P}^{11}$ and its singular locus consists of 27 ordinary double points. This model is used in Section \ref{sec: smoothing} to construct other Calabi--Yau threefolds. The third map was already considered by Kimura in \cite{Ki} and exhibits the birationality of $\mathcal{Z}$ and the variety $V_{3,3}$, a complete intersection of two particular cubics in $\mathbb{P}^5$.\\ The richness of the results obtained from the study of these three maps (such as the geometric relation between rigid Calabi--Yau threefolds and their generalized mirrors or the possibility to construct other Calabi--Yau varieties) suggests it might be interesting to analyze also other models, constructed in a similar way (i.e\ writing the explicit maps defined on $\EEE$). For this reason we introduced maps associated to other divisors (see Remark \ref{rem: hF-kM} and Section \ref{section: other maps}) and in particular we proved that one of these divisors is very ample (and thus gives an isomorphism between $\mathcal{Z}$ and its image under the induced map).\\
We think that our results on the connection between $Pic(\mathcal{Z})$ and a group of matrices and on the geometric relation between $\mathcal{Z}$ and its generalized mirror could lead to a deeper understanding of the generalization of the mirror conjecture to rigid Calabi--Yau threefolds.  \\

The construction of several projective models is important to better understand the geometry of $\mathcal{Z}$, but also because from singular models of $\mathcal{Z}$ we have constructed new Calabi--Yau threefolds.
In sections \ref{sec: other CY 3-folds} and \ref{sec: smoothing} we presented some constructions which can be used to obtain Calabi--Yau threefolds from a given one, in particular from $\mathcal{Z}$. One of these constructions is the smoothing of a singular model of $\mathcal{Z}$. Considering singular models constructed in Section \ref{sec: projective models Z} and applying the smoothing to these models we obtained two distinct  Calabi--Yau threefolds with different Hodge numbers. One of these Calabi--Yau threefolds was unknown until now. It is now clear that the analysis of other singular projective models of $\mathcal{Z}$ (for example associated to the divisors proposed in Section \ref{sec: projective models Z}) could give other unknown Calabi--Yau threefolds. As we said, one can obtain other models considering the maps presented in Section \ref{section: other maps}, but one can also apply the techniques of Section \ref{sec: projective models Z} to other varieties which are desingularizations of quotients of known varieties. This should give a lot of Calabi--Yau threefolds and we hope to find new Calabi--Yau varieties in this way.

\subsection*{Acknowledgements}
{\it The authors are grateful to Bert van Geemen for the idea which underlies this work and for stimulating discussions. We would also like to thank Sergio Cacciatori for several advices and detailed comments on the paper.}

\end{document}